\documentclass[9pt,journal]{IEEEtran}

\usepackage{amsmath}
\usepackage{graphicx}
\usepackage{bm}
\usepackage{cancel}
\usepackage{algorithm}
\usepackage{algorithmicx}
\usepackage{algpseudocode}
\usepackage{epstopdf}
\graphicspath{{../figures}}
\usepackage[font=footnotesize]{subfig}
\usepackage{pbox}
\usepackage{color}
\usepackage{setspace}
\usepackage{booktabs}
\usepackage{cite}
\usepackage{float}
\usepackage{multirow}

\usepackage{color}

\usepackage{amsthm}

\newtheorem{mydef}{Theorem}

\def\Reals{\mbox{\rm I\kern-.2em R}} 

\begin{document}
	%
	\title{Direct  estimation of density functionals using a polynomial basis}
	
	\author{Alan~Wisler,~Visar~Berisha,~Andreas~Spanias,~Alfred O. Hero
		\thanks{This research was supported in part by Office of Naval Research grants N000141410722 (Berisha) and N000141712826 (Berisha) and Army Research Office grant W911NF-15-1- 0479 (Hero).}
	}
	\markboth{}%
	{Shell \MakeLowercase{\textit{et al.}}: Bare Demo of IEEEtran.cls for IEEE Journals}
	\maketitle
	\begin{abstract}
		A number of fundamental quantities in statistical signal processing and information theory can be expressed as integral functions of two probability density functions. Such quantities are called density functionals as they map density functions onto the real line. For example, information divergence functions measure the dissimilarity between two probability density functions and are useful in a number of applications. Typically, estimating these quantities requires complete knowledge of the underlying distribution followed by multi-dimensional integration. Existing methods make parametric assumptions about the data distribution or use non-parametric density estimation followed by high-dimensional integration. In this paper, we propose a new alternative. We introduce the concept of ``data-driven basis functions" - functions of distributions whose value we can estimate given only samples from the underlying distributions without requiring distribution fitting or direct integration. We derive a new data-driven complete basis that is similar to the deterministic Bernstein polynomial basis and develop two methods for performing basis expansions of functionals of two distributions. We also show that the new basis set allows us to approximate functions of distributions as closely as desired. Finally, we evaluate the methodology by developing data driven estimators for the Kullback-Leibler divergences and the Hellinger distance and by constructing empirical estimates of tight bounds on the Bayes error rate.
	\end{abstract}
	\section{Introduction}
	\label{sec:intro}
	
	Information divergence measures play a central role in the fields of machine learning and information theory. Information divergence functions, functionals that map density functions to $\Reals$, have been used in many signal processing applications involving classification \cite{moreno2003kullback}, segmentation \cite{hamza2003image}, source separation \cite{hild2001blind}, clustering \cite{banerjee2005clustering}, and other domains. In machine learning, a sub-class of these divergences known as $f$-divergences \cite{ali1966general}, are widely used as surrogate loss functions since they form convex upper bounds on the non-convex 0-1 loss \cite{nguyen2009surrogate}.
	
	Although these measures prove useful in a variety of applications, the task of estimating them from multivariate probability distributions using finite sample data can pose a significant challenge. It is especially challenging for continuous distributions, which is the focus of this paper. For algorithms and theory for estimating information measures for discrete distributions the reader is referred to \cite{wu2016minimax, jiao2015minimax, wu2015optimal,valiant2011estimating,paninski2003estimation,paninski2004estimating}. For the continuous case treated here, there are  three general classes of methods for estimating divergence \cite{hero01}: 1) parametric methods, 2) non-parametric methods based on density estimation, and 3) non-parametric methods based on direct (or graph-based) estimation. Parametric methods are the most common choice for estimation, and typically offer good convergence rates ($1/N$) when an accurate parametric model is selected. The fundamental limitation of parametric methods, is that an accurate parametric model is rarely available in real world problems and using an inaccurate parametric model can heavily bias the final estimate. As an alternative, when no parametric form is known, non-parametric density estimates such as kernel density estimation \cite{ahmad1976nonparametric}, histogram estimation \cite{gyorfi1987density}, or $k$-nearest neighbor ($k$-NN) density estimation \cite{izenman1991review} are used to characterize the distribution. While these methods are quite powerful in certain scenarios, they are generally high variance, sensitive to outliers, and scale poorly with dimension \cite{hero01}. 
	
	An alternative to these two classes of methods is direct (or graph-based) estimation, which exploits the asymptotic properties of minimal graphs in order to \textit{directly} estimate distribution functionals without ever estimating the underlying distributions themselves. These methods have been used to estimate density functionals such as entropy \cite{kozachenko1987sample,hero1998robust,pal2010estimation}, the information divergence \cite{hero01}, and the $D_p$-divergence \cite{berisha2014empirically}. This class of methods can have faster asymptotic convergence rates \cite{hero01} and are often simpler to implement than plug-in methods which may have many tuning parameters such as kernel width or histogram bin size. Direct estimators are often inspired by an asymptotic property of a graph-theoretic quantity that can be scaled or modified in order to generate an estimator for a given information-theoretic measure. This approach is customized to a specific form of the density functional and is sometimes difficult to generalize. The estimator for the $D_p$ divergence is an example of this approach \cite{berisha2014empirically}. This is in contrast with plug-in estimators, where the same general approach can be used to estimate any distribution functional. Estimators based on influence functions attempt to bridge the gap between the two types of approaches \cite{kandasamy2015nonparametric}. In that work the authors present a recipe for estimating any smooth functional by using a Von Mises expansion - the analog of the Taylor expansion for distributions; however that approach still requires that part of the data be used for density estimation.
	
	In this paper, we provide a general approach for estimating a wide range of distribution functionals. We propose decomposing the functional onto a complete set of ``data-driven" basis functions; where the term ``data-driven" means that the basis is determined directly from the data and does not involve distribution fitting or direct integration. We show that a broad class of distribution functionals can be approximated as closely as desired through linear combinations of our proposed basis, where the weights of the basis expansion are determined through convex optimization. Our approach offers a powerful alternative for estimating information-theoretic distribution functionals. We demonstrate the flexibility of the approach by constructing empirical estimators of bounds on the Bayes error rate using the same basis set. 
	
	The remainder of the paper is organized as follows. In the next section, we review the literature in this area. In Section \ref{sec:Problem} we provide a detailed description of the problem this paper attempts to solve and establish some of the basic mathematical notation used throughout this paper. In Section \ref{sec:basis} we introduce a set of graph-theoretic basis functions and prove that a wide range of information theoretic quantities can be represented by a linear combination of functions in this set. In sections \ref{sec:finite_sample_considerations} and \ref{sec:fitting_routine}, we explore the limitations of the proposed methodology in the finite sample regime and propose two alternate fitting routines to identify weights to map these basis functions to quantities of interest.  In section \ref{sec:divergnece_estimation} we empirically investigate how the proposed method can be used to estimate  popular divergence measures (the KL-divergence, the Hellinger distance, the $D_p$-divergence), and we compare its performance to various parametric and non-parametric alternatives. In Section \ref{sec:bound_estimation}, we show how the method can be extended to form tighter bounds on the Bayes error rate for binary classification problems. Section \ref{sec:conclusion} offers some concluding remarks.
	
	\subsection{Related Work}
	\label{sec:otherwork}
	

		A natural method for non-parametric estimation of continuous distribution functionals involves histogram binning followed by plug-in estimation \cite{silva2010information, darbellay1999estimation}. When the bin-size is adjusted as a function of the number of available samples per bin, this histogram plug-in method is known as Grenander's method of sieves and it enjoys attractive non-parametric convergence rates \cite{geman1982nonparametric,grenander1981abstract}. While these methods may work well for small data dimension ($d=1,2$), their complexity becomes prohibitive for larger dimensions. Recent work has focused on non-parametric plug-in estimators that are more practical in higher dimensions. These approaches generally only estimate the PDF for the values of samples in the reference data, then calculate the expected value across the sample data in place of numerical integration \cite{moon2014multivariate,moon2016improving,poczos2012nonparametric}. To circumvent the slow convergence associated with these approaches, ensemble methods \cite{moon2016improving} and methods based on influence functions \cite{kandasamy2015nonparametric} have been proposed which are capable of achieving the parametric rate $\mathcal{O}(N^{-1})$ MSE convergence if the underlying densities meet certain smoothness conditions \cite{moon2016improving}. Alternatively, estimates of divergence functions that rely on estimates of the likelihood ratio instead of density estimation have been proposed for estimating the $\alpha$-divergence and the $L_2$-divergence \cite{nguyen2009surrogate, nguyen2010estimating, wang2009divergence, wang2009universal, poczos2011estimation}. These methods estimate the likelihood ratio of the two density functions and plug that value into the divergence functions. Other approaches that bypass density estimation are the convex optimization approach of \cite{nguyen2010estimating} to estimate $f$-divergences and the $k$-NN graph and minimal spanning tree  approaches to estimating Henze-Penrose divergence \cite{friedman1979multivariate,henze1999multivariate,berisha2014empirically}.


	Similarly, the approach we propose in this paper bypasses density estimation through a polynomial basis expansion where the basis coefficients are determined through a convex optimization criterion. This provides added flexibility and allows us to easily estimate a large class of distribution functionals and to establish empirical estimates of bounds on the Bayes error rate. Conceptually, this approach is similar to prior work in estimating the entropy of discrete distributions using polynomial approximations \cite{wu2016minimax, jiao2015minimax, wu2015optimal,paninski2003estimation}.

	
	Bounds on optimal performance are a key component in the statistical signal processing literature. For classification problems, it is often desirable to bound the Bayes error rate (BER) - the minimum achievable error in classification problems. The well-known Chernoff upper bound on the probability of error has been used in a number of statistical signal processing applications \cite{chernoff1952measure}. It motivated the Chernoff $\alpha$-divergence \cite{hero01}. The Bhattacharyya distance, a special case of the Chernoff $\alpha$-divergence for $\alpha = \frac{1}{2}$, upper and lower bounds the BER \cite{bhattacharyya1946measure, kailath1967divergence}. Beyond the bounds on the BER based on divergence measures, a number of other bounds exist based on other functionals of the distributions \cite{hashlamoun1994tight, avi1996arbitrarily}. For estimation problems, the Fisher information matrix (FIM) bounds the variance of the optimal unbiased estimator (through it's relationship with the CRLB). The authors have also previously introduced the $D_p$ divergence, a non-parametric $f$-divergence, and showed that it provides provably tighter bounds on the BER than the BC bound \cite{berisha2014empirically}. They extended this work to estimation of the Fisher information in \cite{berisha2014empirical}. 
	
	Our data-driven basis, consisting of Bernstein polynomials, can be used to estimate functionals of distributions and to estimate bounds on Bayes optimal classification performance. Bernstein polynomials of a different form have been used for density estimation \cite{leblanc2012estimating,turnbull2014unimodal,ghosal2001convergence,igarashi2014improving,tenbusch1994two,babu2006smooth}. In contrast to this work, our methods do not rely on density estimation.
	

	\section{Problem Setup}
	\label{sec:Problem}
	In this section, we will set up the problem and establish the notation that will be used throughout the rest of the paper. We are given a set of data $[\mathbf{X},\mathbf{y}]$ containing $N$ instances, where each instance is represented by a $d$-dimensional feature vector $\mathbf{x}_i$ and a binary label $y_i$. Suppose that this data is sampled from underlying distribution, $f_\mathbf{x}(\mathbf{x})$, where
	\begin{equation}
	f_\mathbf{x}(\mathbf{x})=p_0f_0(\mathbf{x})+p_1f_1(\mathbf{x})
	\end{equation}
	is made up of the two conditional class distributions $f_0(\mathbf{x})$ and $f_1(\mathbf{x})$ for classes $0$ and $1$, with prior probabilities $p_0$ and $p_1$ respectively. If the priors aren't explicitly known, they can be easily estimated from the sample data by measuring the ratio of samples drawn from each class. As a simple application of Bayes theorem, we can define the posterior likelihood of class 1, $\eta(\mathbf{x})$, evaluated at a point $\mathbf{x}=\mathbf{x}^*$, as
	\begin{equation}
	\begin{aligned}
	\eta(\mathbf{x}^*)&=P[y=1|\mathbf{x}=\mathbf{x}^*]=\frac{P[y=1]f_{\mathbf{x}}(\mathbf{x}^*|y=1)}{f_{\mathbf{x}}(\mathbf{x}^*)}\\
	&=\frac{p_1f_1(\mathbf{x}^*)}{f_{\mathbf{x}}(\mathbf{x}^*)}=\frac{p_1f_1(\mathbf{x}^*)}{p_0f_0(\mathbf{x}^*)+p_1f_1(\mathbf{x}^*)}
	\label{eq:posterior}
	\end{aligned}
	\end{equation} 
	We can similarly define the posterior probability for class 0 as
	\begin{equation}
	P[y=0|\mathbf{x}=\mathbf{x}^*]=\frac{p_0f_0(\mathbf{x}^*)}{p_0f_0(\mathbf{x}^*)+p_1f_1(\mathbf{x}^*)},
	\end{equation}
	and since $y$ is binary,
	\begin{equation}
	P[y=0|\mathbf{x}=\mathbf{x}^*]=1-\eta(\mathbf{x}^*).
	\end{equation}
	To simplify the notation, we remove the dependence of $\eta$ on $\mathbf{x}^*$ from portions of the analysis that follow. 
	
	Suppose that we wish to estimate some functional $G(f_0,f_1)$ of distributions $f_0(\mathbf{x})$ and $f_1(\mathbf{x})$, which can be expressed in the following form
	\begin{equation}
	G(f_0,f_1)=\int g(\eta(\mathbf{x}))f_{\mathbf{x}}(\mathbf{x}) d\mathbf{x}.
	\label{eq:form}
	\end{equation}
	Throughout the rest of this paper, we will refer to the $g(\eta)$ in (\ref{eq:form}) as the posterior mapping function. Many functionals in machine learning and information theory, such as $f$-divergences and loss functions, can be expressed this way. Consider the family of $f$-divergences as an example. They are defined as
	\begin{equation}
	D_{\phi}(f_0,f_1)=\int \phi\bigg(\frac{f_0(\mathbf{x})}{f_1(\mathbf{x})}\bigg)f_1(\mathbf{x})d\mathbf{x},
	\label{eq:fdiv1}
	\end{equation}
	where $\phi(t)$ is a convex or concave function unique to the given $f$-divergence. By substituting 
	\begin{equation}
	\frac{f_0(\mathbf{x})}{f_1(\mathbf{x})}=\frac{p_1(1-\eta(\mathbf{x}))}{p_0\eta(\mathbf{x})}
	\end{equation}
	and
	\begin{equation}
	f_1(\mathbf{x})=\frac{\eta(\mathbf{x})}{p_1}f_{\mathbf{x}}(\mathbf{x})
	\end{equation}
	we can redefine (\ref{eq:fdiv1}) as
	\begin{equation}
	D_{\phi}(f_0,f_1)=\int \phi\bigg(\frac{p_1(1-\eta(\mathbf{x}))}{p_0\eta(\mathbf{x})}\bigg)\frac{\eta(\mathbf{x})}{p_1}f_{\mathbf{x}}(\mathbf{x})d\mathbf{x}.
	\end{equation}
	Thus any $f$-divergence can be presented in the form outlined in \eqref{eq:form} simply by defining the posterior mapping function $g(\eta)$ as
	\begin{equation}
	g(\eta)=\frac{\eta}{p_1}\phi\bigg(\frac{p_1(1-\eta)}{p_0\eta}\bigg).
	\end{equation}
	We propose a procedure for estimating these types of divergence functionals which bypasses density estimation. We do this by representing the functional in terms of the asymptotic limit of a linear combination of graph-theoretic basis functions.
	
	Suppose that there exists a set of basis functions $H_0(\eta),...,H_k(\eta)$ that can be similarly expressed as
	\begin{equation}
	H_i(f_0,f_1)=\int h_i(\eta(\mathbf{x}))f_{\mathbf{x}}(\mathbf{x}) d\mathbf{x}.
	\label{eq:basis1}
	\end{equation}
	If we assume that there exists a set of coefficients such that 
	
	\begin{equation}
	g(\eta)\approx \sum_{i=0}^{k}w_ih_i(\eta),
	\end{equation}
	then consequently
	\begin{equation}
	G(f_0,f_1)\approx \hat{G}(f_0,f_1)= \sum_{i=0}^{k}w_iH_i(\eta),
	\end{equation}
	where the sense of approximation is that the $\ell_2$ norm of the difference between the right and left hand sides is small. In the following section, we will introduce a set of basis functions that have the desired properties.
	
	\section{Graph-theoretic Basis Functions}
	\label{sec:basis}
	\begin{figure}[tb]
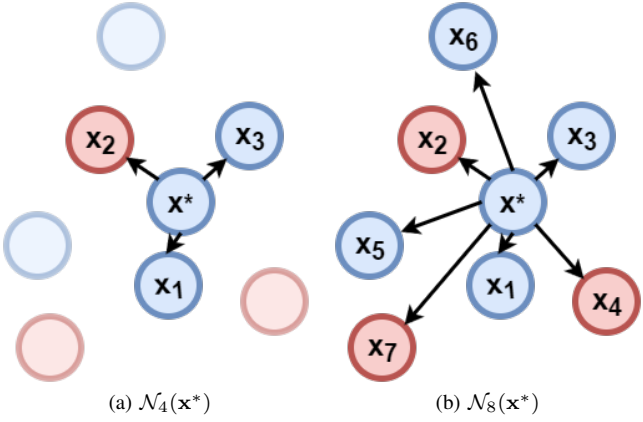

		\begin{center}
			\subfloat[$\mathcal{N}_4(\mathbf{x}^*)$]{
				\includegraphics[width=0.225\textwidth]{NN_Diagram1.png}
				\label{fig:NNex1}
			}
			\subfloat[$\mathcal{N}_8(\mathbf{x}^*)$]{
				\includegraphics[width=0.225\textwidth]{NN_Diagram2.png}
				\label{fig:NNex2}
			}
			\caption{Illustration of two neighborhoods of $\mathbf{x}^*$ for $k=4$ and $k=8$, instances with $y=0$ are blue while instances with $y=1$ are red. In the first scenario $\Phi_4(\mathbf{x}^*)=1$, since only one instance in $\mathcal{N}_4(\mathbf{x}^*)$ is red. In the second scenario $\Phi_8(\mathbf{x}^*)=3$, since three of the eight instance in $\mathcal{N}_8(\mathbf{x}^*)$ are red.}
			\label{fig:Neighborhood illustrations}
		\end{center}
	\end{figure}

	Consider the dataset $[\mathbf{X},\mathbf{y}]$ previously defined. Suppose we select an arbitrary instance $\mathbf{x}^*$ from $\mathbf{X}$ and examine it along with the set of its $k-1$ nearest neighbors $\mathbf{x}_{NN}^1,\mathbf{x}_{NN}^2,...,\mathbf{x}_{NN}^{k-1}$ in $\mathbf{X}$. We can define the neighborhood set $\mathcal{N}_k(\mathbf{x}^*)=[\mathbf{x}^*,\mathbf{x}_{NN}^1,...,\mathbf{x}_{NN}^{k-1}]$, as the union of $\mathbf{x}^*$ and its $k-1$ nearest neighbors. Using this we define $\Phi_k(\mathbf{x}^*)$ as the number of instances in the neighborhood set which are drawn from class 1, or alternatively, the sum of $y$ across all points in the neighborhood  
	\begin{equation}
	\Phi_k(\mathbf{x}^*)=\sum_{i: \mathbf{x}_i\in\mathcal{N}_k(\mathbf{x}^*)}^{}y_i.
	\end{equation} 
	
	Figure \ref{fig:Neighborhood illustrations} provides a simple illustration to help explain how $\Phi_k(\mathbf{x}^*)$ is calculated. Calculating $\Phi_k$ is similar to how nearest neighbor classifiers make decisions, but with two important differences:
	\begin{enumerate}
		\item The base instance $\mathbf{x}^*$ is considered in the neighborhood indistinguishably from other instances in $\mathcal{N}_k(\mathbf{x}^*)$.
		\item Where traditional $k$-NN classifiers are concerned only with identifying the majority, we are interested in the exact number of instances drawn from each class.
	\end{enumerate}
	In essence $\Phi_k(\mathbf{x}^*)$ tells us something about the probability that $y=1$ for instances on or near $\mathbf{x}^*$. Since we are more concerned with the dataset as a whole than the local characteristics in $\mathbf{x}$, we define the statistic $\rho_{r,k,N}(\mathbf{X})$ to be the fraction of instances $\mathbf{x}\in \mathbf{X}$, for which $\Phi_k(\mathbf{x})=r$, $r\leq k$. If we define the indicator function $I_{r,k}(\mathbf{x})$ as
	\begin{equation}
	I_{r,k}(\mathbf{x})=\begin{cases} 
	1 & \Phi_k(\mathbf{x})=r\\
	0 & otherwise,
	\end{cases}
	\end{equation}
	then this test statistic $\rho_{r,k,N}(\mathbf{X})$ can be represented by 
	\begin{equation} \label{eqn:rhostat}
	\rho_{r,k,N}(\mathbf{X})=\frac{1}{N}\sum_{\mathbf{x}\in \mathbf{X}}^{}I_{r,k}(\mathbf{x}).
	\end{equation}
	The function $\rho_{r,k,N}(\mathbf{X})$ is simply the proportion of $k$-NN neighborhoods that contain exactly $r$ points from class $y=1$. This statistic has a number of desirable qualities. We show that  this statistic asymptotically converges to a function of the underlying distributions that can be described in the form outlined in (\ref{eq:form}). The following is proven in Appendix \ref{sec:proof_thm1}.

	\begin{mydef} 
		As the number of samples ($N$) approaches infinity, 
		\begin{equation*}
		\lim\limits_{N\rightarrow \infty }\rho_{r,k,N}(\mathbf{X})\overset{L^2}{\to}\int \dbinom{k}{r}\eta^r(\mathbf{x})(1-\eta(\mathbf{x}))^{k-r}f_{\mathbf{x}}(\mathbf{x})d\mathbf{x} 
		\end{equation*}
		whenever $k/N\rightarrow 0$.
		\label{thm:Basis1}
	\end{mydef}

	We propose to use the asymptotic form of $\rho_{r,k,N}(\mathbf{X})$ defined in Theorem \ref{thm:Basis1} as a basis function for estimating functionals of the form \eqref{eq:form},
	\begin{align}
	H_{r,k}(f_0,f_1) & = \lim\limits_{N\rightarrow \infty}\rho_{r,k,N}(\mathbf{X}) 
	& =  \int h_{r,k}(\eta(\mathbf{x})) f_{\mathbf{x}}(\mathbf{x})d\mathbf{x}
	\end{align}
	where
	\begin{equation}
	h_{r,k}(\eta)=\dbinom{k}{r}\eta^r(1-\eta)^{k-r}.
	\label{eq:bernstein_basis}
	\end{equation}
	
	The function \eqref{eq:bernstein_basis} is the $r^{th}$ Bernstein basis polynomial of degree $k$ \cite{lorentz2012bernstein}. Bernstein's proof of the Weierstrass Approximation Theorem \cite{bernstein1912démo} asserts that any continuous function $g(\eta)$ can be uniformly approximated on $\eta\in [0,1]$ to any desired accuracy by a linear combination of functions in \eqref{eq:bernstein_basis} of the form
	\begin{equation}
	g(\eta)\approx \sum_{r=0}^k g\Big(\frac{r}{k}\Big)   h_{r,k}(\eta).
	\end{equation} 
	
	Combining this result with Theorem \ref{thm:Basis1}, we can show that a linear combination of this basis can be used to estimate any function of the form (\ref{eq:form}) .
	
	\begin{mydef} For any $G(f_0,f_1)$ that can be expressed in the form
		\begin{equation*}
		G(f_0,f_1)=\int g(\eta(\mathbf{x}))f_{\mathbf{x}}(\mathbf{x})d\mathbf{x},
		\end{equation*}
		where $g(\eta)$ is continuous on $[0,1]$, the approximation 
		\begin{equation}
		\hat{G}_{k,N}(\mathbf{X})=\sum_{r=0}^{k}g\Big(\frac{r}{k}\Big)\rho_{r,k,N}(\mathbf{X})
		\label{eq:bernstein_sum}
		\end{equation}
		satisfies
		\begin{equation}
	\lim\limits_{k\rightarrow \infty}\lim_{\substack{N\to \infty\\ k/N\to 0}} E\bigg[\Big( \hat{G}_{k,N}(\mathbf{X})-G(f_0,f_1)\Big)^2\bigg]=0.
	\end{equation}
		\label{thm:final}
	\end{mydef}

	Theorem \ref{thm:final} provides an asymptotically consistent method of estimating a variety of information-theoretic functions that makes no assumptions on the underlying distributions and can be calculated without having to perform density estimation. Throughout the rest of the paper we will refer to the weights $g(r/k)$ in the approximation specified by \eqref{eq:bernstein_sum} in Theorem \ref{thm:final}   as the Bernstein weights. We next turn to the finite sample properties of the estimator $\rho_{r,k,N}(\mathbf{X})$.
	\section{Finite Sample Considerations}
	\label{sec:finite_sample_considerations}
	
	The previous Section investigated the asymptotic properties of linear combinations of the proposed set of empirically estimable basis functions. The asymptotic consistency of the proposed method is valuable, however in real world scenarios, data is inherently a finite resource, and as a result the efficacy of this method is heavily dependent on its convergence characteristics in the finite sample regime. In this section, we will take a detailed look into how restricting both $N$ and $k$ affects our ability to estimate functions of two distributions. To do this, it is necessary to first break down the different sources of error in the proposed methodology.
	
	\subsection{Estimation vs. Approximation Error}
	\label{ssec:errortypes}
	The goal of this paper is to empirically estimate the functional $G(f_0,f_1)$ of the two underlying distributions $f_0(\mathbf{x})$ and $f_1(\mathbf{x})$ using a linear combination of directly estimable basis functions 
	\begin{equation}
	\hat{G}_{k,N}(\mathbf{X})=\sum_{r=0}^{k}w_r\hat{H}_{r,k,N}(\mathbf{X}).
	\end{equation}
	We divide the error of this estimate into two types, the approximation error ($e_A$) and the estimation error ($e_{est}$):
	\begin{equation}
	\begin{aligned}
	e_T&=G(f_0,f_1)-\hat{G}_{k,N}(\mathbf{X})\\
	&= \underbrace{G(f_0,f_1)-\hat{G}_k(f_0,f_1)}_{=:e_A}+\underbrace{\hat{G}_k(f_0,f_1)-\hat{G}_{k,N}(\mathbf{X})}_{=:e_{est}}
	.
	\end{aligned}
	\label{eq:err1}
	\end{equation}
	This allows us to separate the error in estimating the basis functions from error in fitting to the posterior mapping function. Understanding the trade-off between these two error types will be particularly useful in Section \ref{sec:fitting_routine}, where we explore different methods of fitting weights to the desired density functionals.

	\subsection{Considerations for finite $k$}
	A finite sample also implies a finite $k$ and impacts the approximation error. Let us consider the Bernstein weighting scheme introduced in \eqref{eq:bernstein_sum} for the scenario where the size of the basis set ($k$) is restricted. Consider the following example problem.
	
	\vspace{0.2cm}
	\noindent \textit{Example:} Suppose that we wish to estimate the function 
	\begin{equation}
	g(\eta)=\dbinom{3}{1}\eta(1-\eta)^2
	\end{equation}
	using the basis set $\beta_{0,3}(\eta),\beta_{1,3}(\eta),\beta_{2,3}(\eta),\beta_{3,3}(\eta)$. Because $g(\eta)=\beta_{1,3}(\eta)$, there exists a set of weights such that 
	\begin{equation}
	\sum_{r=0}^{3}w_r \beta_{r,3}(\eta)=g(\eta),
	\end{equation}
	however, using the Bernstein weighting scheme in \eqref{eq:bernstein_sum} yields
	\begin{equation}
	\begin{aligned}
	\hat{g}(\eta)=&\sum_{r=0}^{3}g\Big(\frac{r}{3}\Big) \beta_{r,3}(\eta)\\
	=&g\Big(\frac{0}{3}\Big) \beta_{0,3}(\eta)+g\Big(\frac{1}{3}\Big) \beta_{1,3}(\eta)+g\Big(\frac{2}{3}\Big) \beta_{2,3}(\eta)\\&+g\Big(\frac{3}{3}\Big) \beta_{3,3}(\eta)\\
	=&\frac{4}{3}\eta(1-\eta)^2+\frac{2}{3}\eta^2(1-\eta)\\ 
	\neq &g(\eta).
	\end{aligned}
	\end{equation}
	
	It is clear from this example that the Bernstein weighting procedure do not always provide ideal weights when $k$ is restricted. Based on these results, we are motivated to explore alternative weighting procedures in order to improve the performance of this method for the finite sample case. In the following Section, we will introduce a method of finding better weights using convex optimization.

	\begin{figure*}[!tb]
		\begin{center}
			\includegraphics[width=0.99\textwidth]{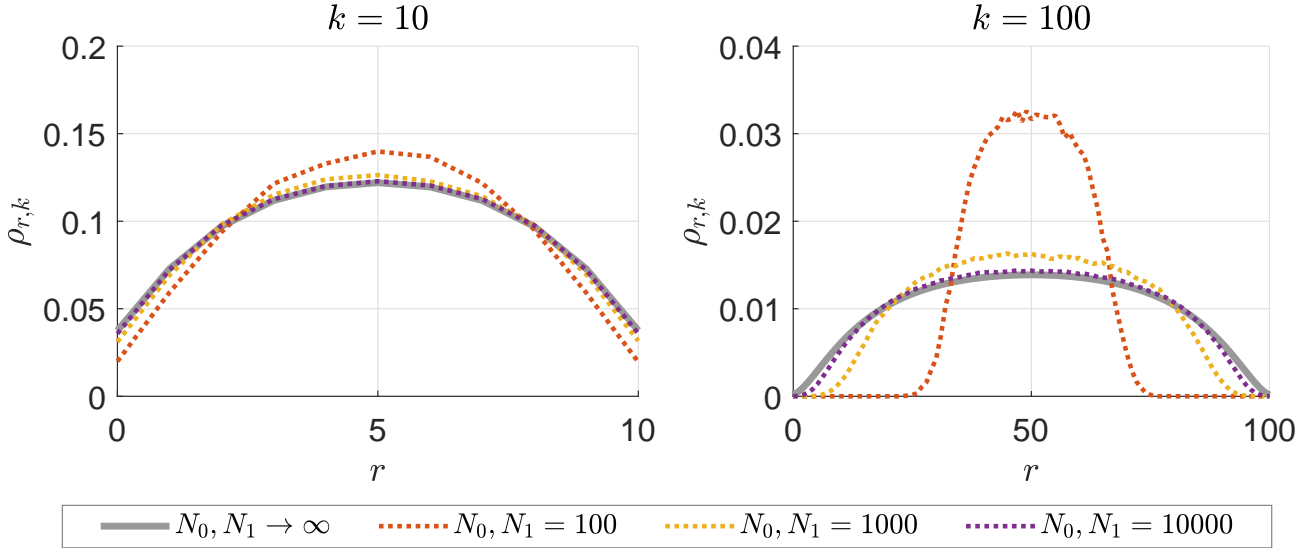}
			\caption{Plot of true and estimated basis values vs. $r$ for data drawn from underlying distributions $f_0(\mathbf{x})\sim \mathcal{N}(\mathbf{0}_3,\mathbf{I}_3)$ and $f_1(\mathbf{x})\sim \mathcal{N}(\frac{1}{\sqrt{3}}\mathbf{1}_3,\mathbf{I}_3)$.}
			\label{fig:basisplots}
		\end{center}
	\end{figure*}

	Regardless of how weights are assigned to the approximation, selection of $k$ remains an important factor affecting performance. In order to satisfy Theorem \ref{thm:final}, $k$ should be functionally dependent on $N$, such that $N$ approaches infinity, both $k\rightarrow \infty$ and $\frac{k}{N}\rightarrow 0$, however this still provides a great degree of freedom in the selection of $k$. 

	In general, there are two major competing factors that must be considered when selecting $k$. The first is that the Weierstrass approximation theorem can exactly represent any posterior mapping function $g(\eta)$ as a linear combination of the proposed basis set only as $k\rightarrow\infty$. This provides motivation for choosing a large $k$-value to ensure the best possible fit of $g(\eta)$. The second factor is that the asymptotic characteristics of $\rho_{r,k,N}(\mathbf{X})$ are dependent on all points in $\mathcal{N}_k(\mathbf{x}^*)$. Moreover the regime for which Theorem \ref{thm:final} holds, requires that $\frac{k}{N}\rightarrow 0$, so we are motivated to select $k$ such that $N \gg k$. This means that the selection of $k$ must achieve a compromise in the trade-off between the approximation and estimation errors, since larger values of $k$ will increase the amount of finite sample error made in estimating the individual basis functions, while lower values of $k$ may inhibit our ability to accurately model the desired function in the asymptotic regime. 
	
	To illustrate how our ability to estimate the desired set of basis functions varies with $k$, we calculate the estimated and asymptotic values of $\rho_{r,k,N}(\mathbf{X})$ for $k=10$ and $k=100$ on data drawn from distributions $f_0(\mathbf{x})\sim \mathcal{N}(\mathbf{0}_3,\mathbf{I}_3)$ and $f_1(\mathbf{x})\sim \mathcal{N}(\frac{1}{\sqrt{3}}\mathbf{1}_3,\mathbf{I}_3)$, where $\mathbf{0}_3=[0 \ \  0 \ \  0]$ and $\mathbf{1}_3=[1 \ \  1 \ \  1]$, and plot the results in Figure \ref{fig:basisplots}. Estimates are calculated at 3 different sample sizes ($N_0=N_1=100,1000,10000$) and each estimate shown in Figure \ref{fig:basisplots} has been averaged across 500 Monte Carlo trials. While we can estimate the basis set for either $k$-value with a high degree of accuracy given enough samples, the estimates for $k=10$ are noticeably more accurate. In fact, we are able to do about as well with $1000$ samples for $k=10$ as we are with $10000$ samples for $k=100$.
	
	\section{Optimization Criteria for Fitting Density Functionals}
	\label{sec:fitting_routine}
	In this section, we propose a convex optimization criterion to identify appropriate weights for fitting information-theoretic functions when $k$ and $N$ are restricted. Inherently, our goal is to minimize the total error, defined in (\ref{eq:err1}), however minimizing this quantity directly isn't feasible since the value of $G(f_0,f_1)$ is unknown. To circumvent this challenge we focus on developing a criterion to minimize the approximation error.  We initially develop an optimization criterion that assumes the posterior is uniformly distributed, then propose an alternate method which incorporates an estimate of the posterior density function in order to more accurately model the approximation error.
	\subsection{Uniform Optimization Criteria}
	\label{ssec:general_fitting_routine}
	Recall that the approximation error $e_A$ can be represented as
	\begin{equation}
	e_A=\int \epsilon(\eta(\mathbf{x}))f_\mathbf{x}(\mathbf{x})d\mathbf{x},
	\label{eq:eA1}
	\end{equation}
	where 
	\begin{equation}
	\epsilon(\eta)=g(\eta)-\sum_{r=0}^{k}w_rh_{r,k}(\eta).
	\end{equation}
	Since solving (\ref{eq:eA1}) requires high-dimensional integration and knowledge of the underlying distributions. However, because $\eta$ is a function of $\mathbf{x}$, $\epsilon(\eta)$ is implicitly a function of $\mathbf{x}$ as well, and by the law of the unconscious statistician \cite{gubner2006probability},
	\begin{equation}
	e_A=E[\epsilon(\eta)]=\int \epsilon(\eta)f_{\eta}(\eta)d\eta,
	\label{eq:eAfinal}
	\end{equation} 
	where $f_{\eta}(\eta)$ is the probability density function of the random variable $\eta$. Rewriting the error in this form simplifies the region of integration to a well defined space (since $\eta\in[0,1]$) and circumvents the high dimensionality of $\mathbf{x}$. While this eliminates some of the challenges in calculating the error it also creates new ones stemming from the fact that $f_{\eta}(\eta)$ is unknown and difficult to estimate due to its implicit dependency on $f_0(\mathbf{x})$ and $f_1(\mathbf{x})$. The task of estimating $f_{\eta}(\eta)$ will be explored in detail in Section \ref{sec:post_dist_estimation}, however for the time being we will bypass this challenge and simply attempt to minimize
	\begin{equation}
	e_A^*=\int |\epsilon(\eta)|^2 d\eta.
	\label{eq:eAstar}
	\end{equation} 
	It is worth noting that if $f_{\eta}(\eta)$ is uniformly distributed
	\begin{equation}
	e_A^*=E\big[|\epsilon(\eta)|^2\big] \geq e_A^2.
	\end{equation}
	While there exists an analytical solution for identifying the weights which minimize \eqref{eq:eAstar}\cite{farouki2000legendre}, we use a convex optimization procedure that allows us to also account for the estimation error. If we define a discretized set of posterior values $\boldsymbol{\tilde{\eta}}=[\tilde{\eta}_1,\tilde{\eta}_2,...,\tilde{\eta}_{\tilde{N}}]$, where $0 \leq \tilde{\eta}_1 < \tilde{\eta}_2 <...<\eta_{\tilde{N}} \leq 1$, a procedure to identify weights that minimize \eqref{eq:eAstar} can be defined as
	\begin{equation}
	w_0,...,w_k=\underset{w_0,...,w_k}{\mathrm{argmin}}\frac{1}{\tilde{N}}\sum_{i=1}^{\tilde{N}}\Big | g(\tilde{\eta}_i)-\sum_{r=0}^{k}w_r h_{r,k}(\tilde{\eta}_i)\Big |^2.
	\label{eq:fitting_procedure1}
	\end{equation}
	. 
	
\begin{figure}[t]
		\includegraphics[width=0.46\textwidth]{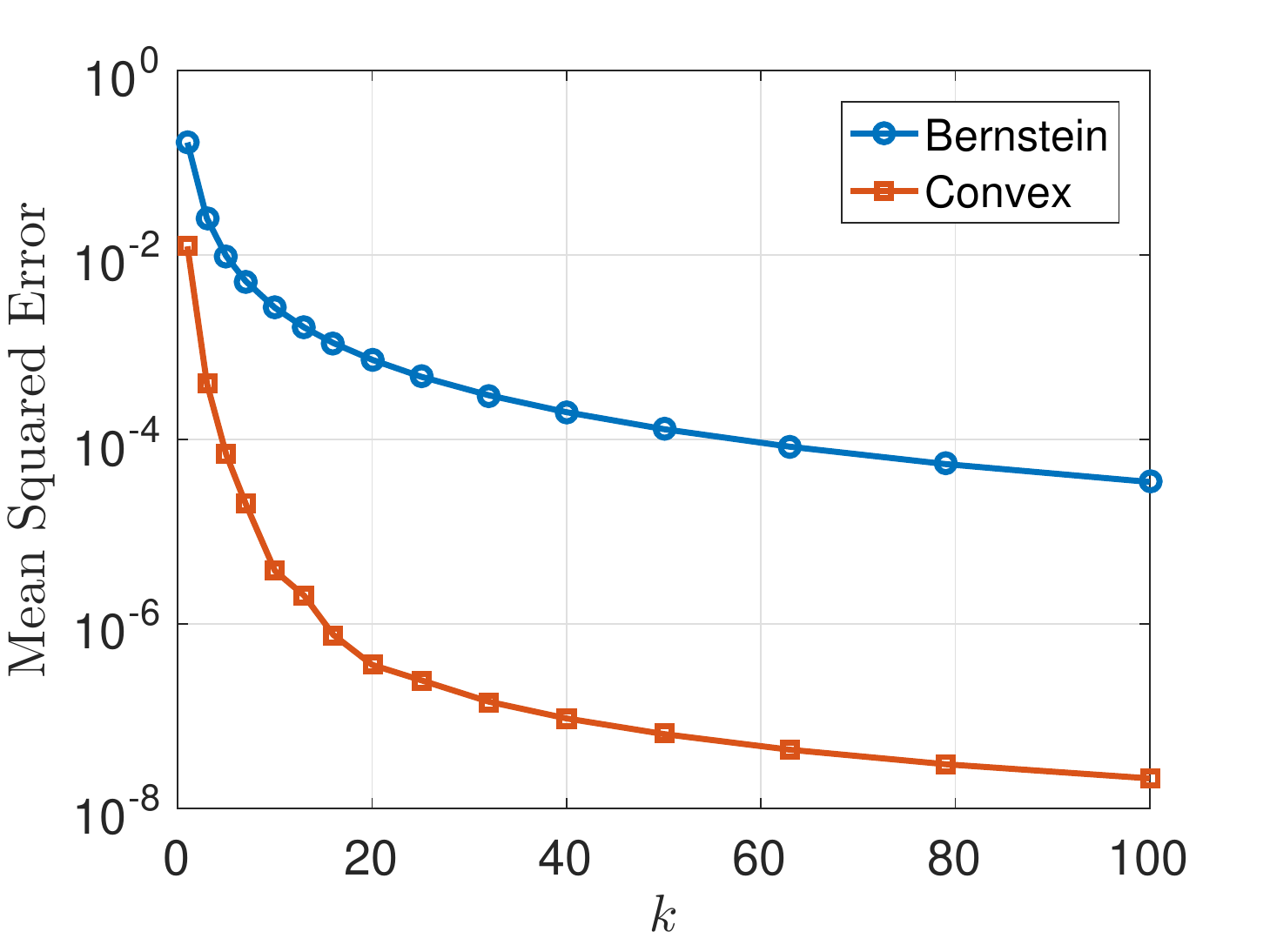}
		\caption{Approximation error of each fitting procedure as a function of the number of basis elements $k$. This is the idealized case where the basis estimation error is zero and the total error is solely due to imperfect approximation of the posterior mapping function $g(\eta)$.}
		\label{fig:bernstein_kplot}
	\end{figure}

	To illustrate the effectiveness of this method, we consider the example problem of trying to estimate the Hellinger distance (a problem we will further explore in Section \ref{sec:divergnece_estimation}). If we assume both classes have equal prior probability ($p_0=p_1=0.5$), then the posterior mapping function for the squared Hellinger distance is 
	\begin{equation}
	g(\eta)=(\sqrt{\eta}-\sqrt{1-\eta})^2.
	\end{equation}
	This function is estimated using this convex weighting procedure as well as the previously described Bernstein weighting procedure, and we compare how well each method models the desired function for values of $k$ varying from $0$ to $100$. The performances of each method is evaluated by the following formula
	\begin{equation}
	\mathrm{MSE}(\hat{g},g)=\sum_{i=1}^{\tilde{N}}\Big | g(\tilde{\eta}_i)-\hat{g}(\tilde{\eta}_i)\Big |^2,
	\label{eq:mse_g}
	\end{equation}
	and the results are presented in Figure \ref{fig:bernstein_kplot} for a range of $k$ values varying from $1$ to $100$. This experiment shows that the proposed convex fitting procedure is able to approximate the Hellinger posterior mapping function far more accurately than the Bernstein approximation. This improvement isn't surprising since the proposed method directly minimizes the MSE whereas the Bernstein approximation guarantees consistency only as $k\rightarrow \infty$, but it still helps to illustrate the potential for improvement in the Bernstein weights that exists for smaller $k$.
	
\begin{figure}[t]
		\includegraphics[width=0.5\textwidth]{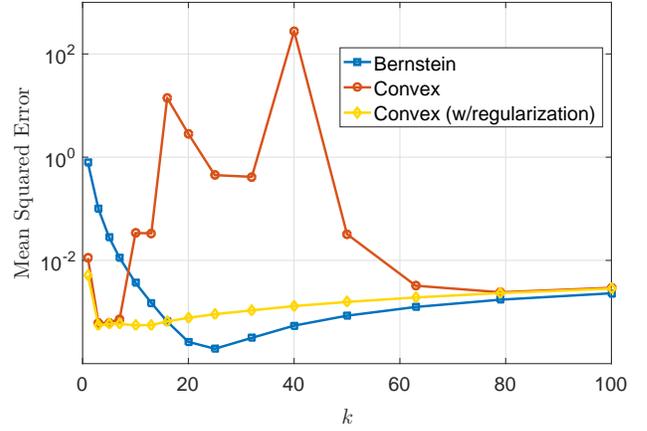}
		\caption{The total error of each fitting procedure as a function of $k$, when there is both approximation and estimation error.}
		\label{fig:bernstein_kplot2}
	\end{figure}

	The expression \eqref{eq:mse_g}  does not take into account finite sample errors that lead to noisy estimates of the basis functions and thus does not directly reflect our ability to estimate $G(f_0,f_1)$ with a finite sample. The consequences of this could be quite significant. When using a similar approach for entropy estimation, Paninski observed that the good approximation properties were a result of ``large oscillating coefficients'' which magnify the variance of the corresponding estimator\cite{paninski2003estimation}. Additionally, it does not account for the possibility that $\eta$ is distributed non-uniformly.
	To empirically examine the finite sample properties of the two approaches, we repeat the previous experiment, this time estimating the basis functions empirically on samples of data drawn from distributions $f_0(\mathbf{x})\sim \mathcal{N}( \mathbf{0}_3,\mathbf{I}_3)$ and $f_1(\mathbf{x})\sim \mathcal{N}( \frac{\mathbf{1}_3}{\sqrt{3}},\mathbf{I}_3)$. We generate $N=1000$ samples (500 samples per class) in each of the 500 iterations of a Monte Carlo simulation, and evaluate the MSE as
	\begin{equation}
	\mathrm{MSE}(G, \hat{G})=\frac{1}{N_{MC}}\sum_{i=1}^{N_{MC}}\big[ G(f_0,f_1)-\hat{G}(f_0,f_1)\big]^2,
	\end{equation}
	where $N_{MC}$ represents the number of Monte Carlo iterations. Since we know that the estimation error is scaled by the magnitude of the weights, we also evaluate a modified fitting routine which augments \eqref{eq:fitting_procedure1} with a regularization term to penalize solutions with large weights,
	
	\begin{equation}
	\begin{aligned}
	&w_0,...,w_k=\\ &\underset{w_0,...,w_k}{\mathrm{argmin}}\frac{1}{\tilde{N}}\sum_{i=1}^{\tilde{N}}\Big | g(\tilde{\eta}_i)-\sum_{r=0}^{k}w_r h_{r,k}(\tilde{\eta}_i)\Big |^2
	+\frac{\lambda}{k}\sum_{r=0}^{k}w_r^2,
	\end{aligned}
	\label{eq:fitting_procedure_wRegularization}
	\end{equation}
	where $\lambda$ represents a tuning parameter which controls the importance assigned to minimization of the approximation error relative to the estimation error. Intuitively, higher $\lambda$ values will make sense for smaller data sets to control the variance of the estimator. We set $\lambda=0.01$ for all experiments conducted in this paper.
	The results of this experiment are shown in Figure \ref{fig:bernstein_kplot2}. We immediately see the necessity of the regularization term, as without it the error becomes extremely large for a range of $k$ values. More generally, the inclusion of the regularization term improves the performance at every $k$ value in this experiment. In comparing the Bernstein weights with the convex (regularized) weights, we find that 1) the performance of the convex method is less dependent on the selection of $k$ and 2) the convex weights generally perform better at lower values of $k$, while the Bernstein weights outperform at higher $k$. While the peak performance of the Bernstein method is higher than the convex method, there exists no good method of selecting $k$ {\em a priori} in order to reliably achieve this performance.  In contrast, the convex method with regularization is less sensitive to the value of $k$ selected.

	\subsection{Density-weighted Optimization Criteria}
	\label{sec:post_dist_estimation}
	
	In the optimization criteria introduced in the previous section we implicitly make the assumption that the distribution of the random variable $\eta(\mathbf{x}) \sim f_{\eta}(\eta)$ is uniformly distributed. In this section we will investigate a data-driven estimator for  $f_{\eta}(\eta)$. However, before we proceed it is important to clarify what this distribution actually is. 
	
	\begin{figure*}[!t]
		
		\begin{center}
			\subfloat[Class distribution]{
				\includegraphics[width=0.33\textwidth]{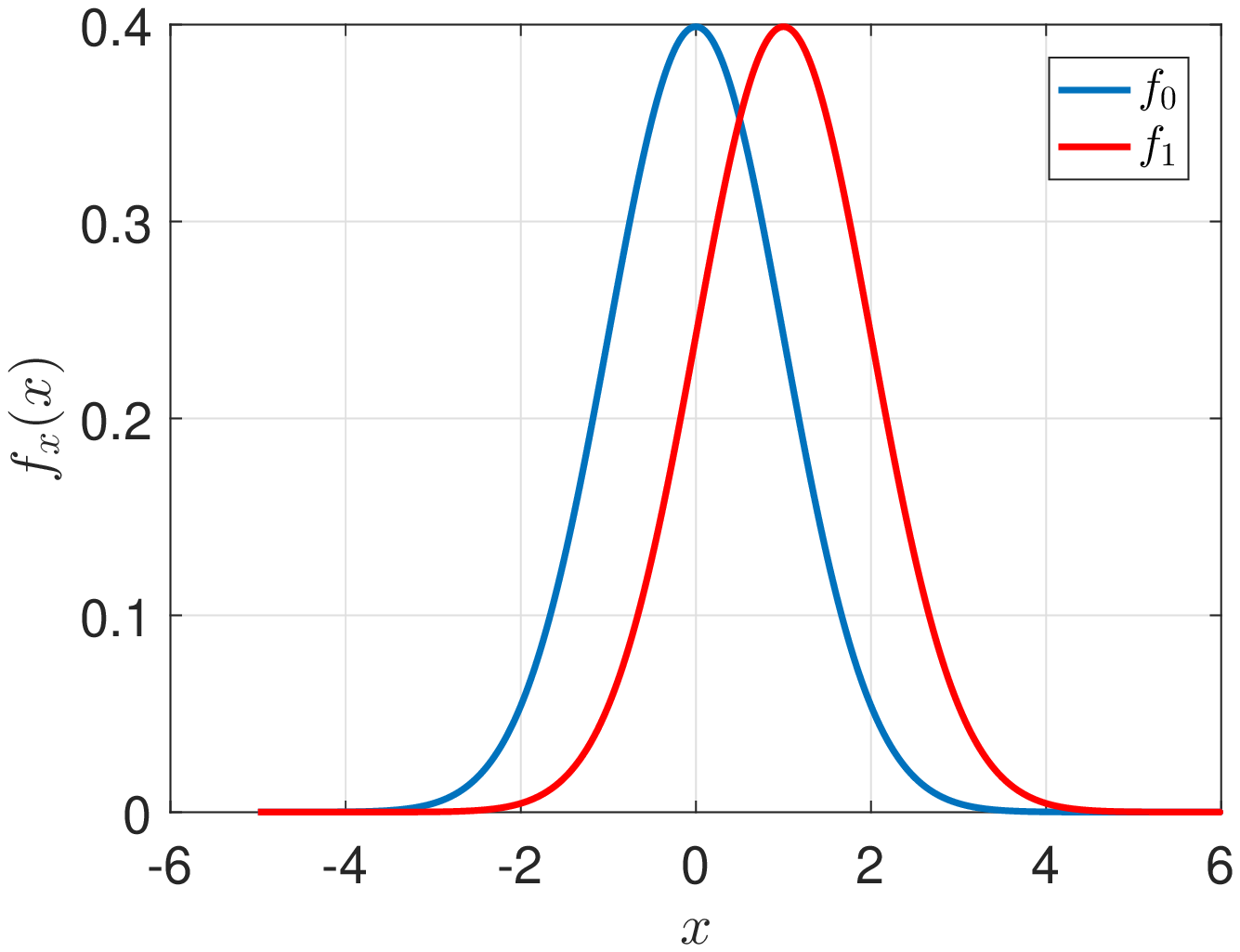}
				\label{fig:f_plot_sep1}
			}	
			\subfloat[Class 1 Posterior]{		
				\includegraphics[width=0.33\textwidth]{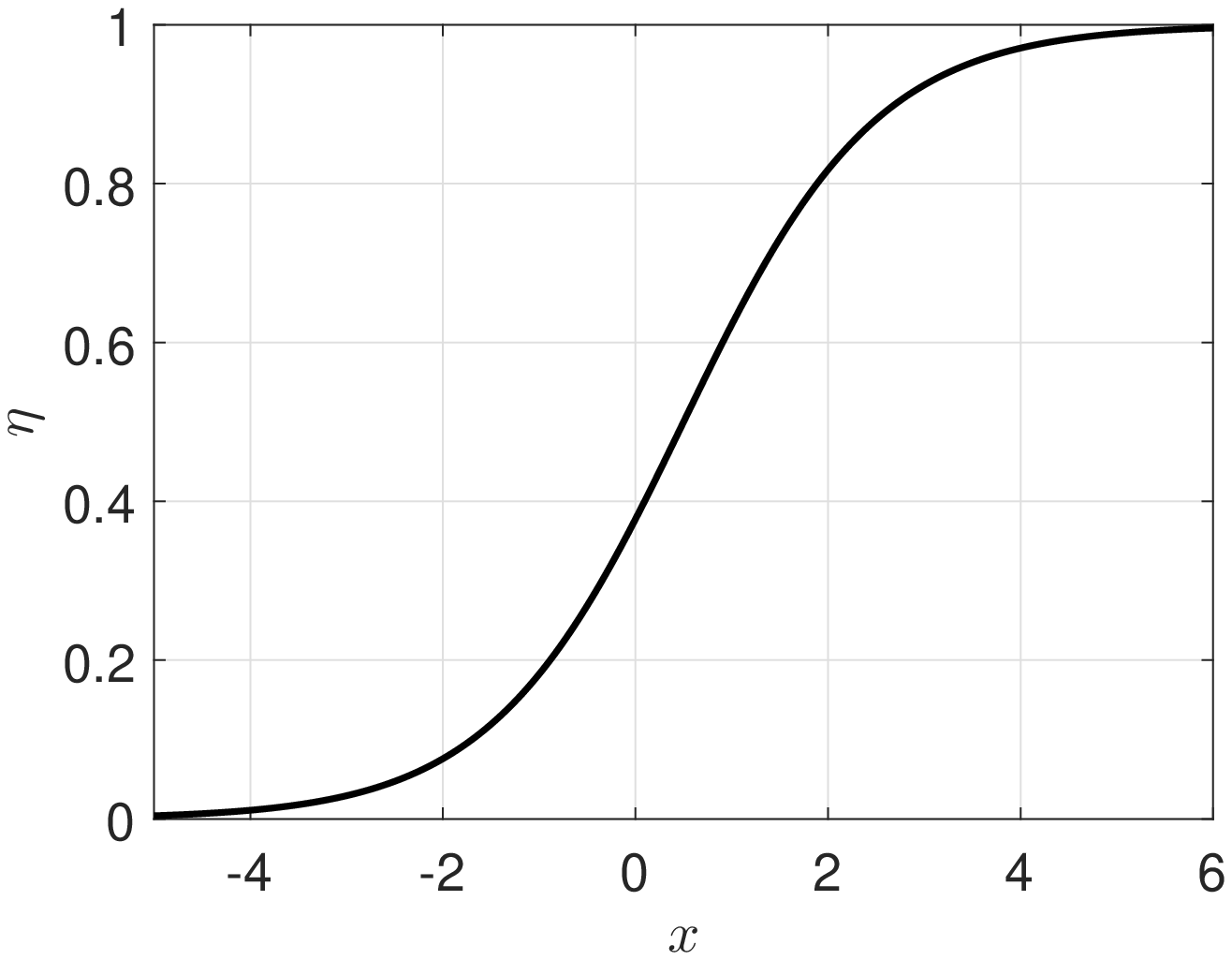}
				\label{fig:eta_plot_sep1}
			}
			\subfloat[Distribution of class 1 posterior likelihoods]{
				\includegraphics[width=0.33\textwidth]{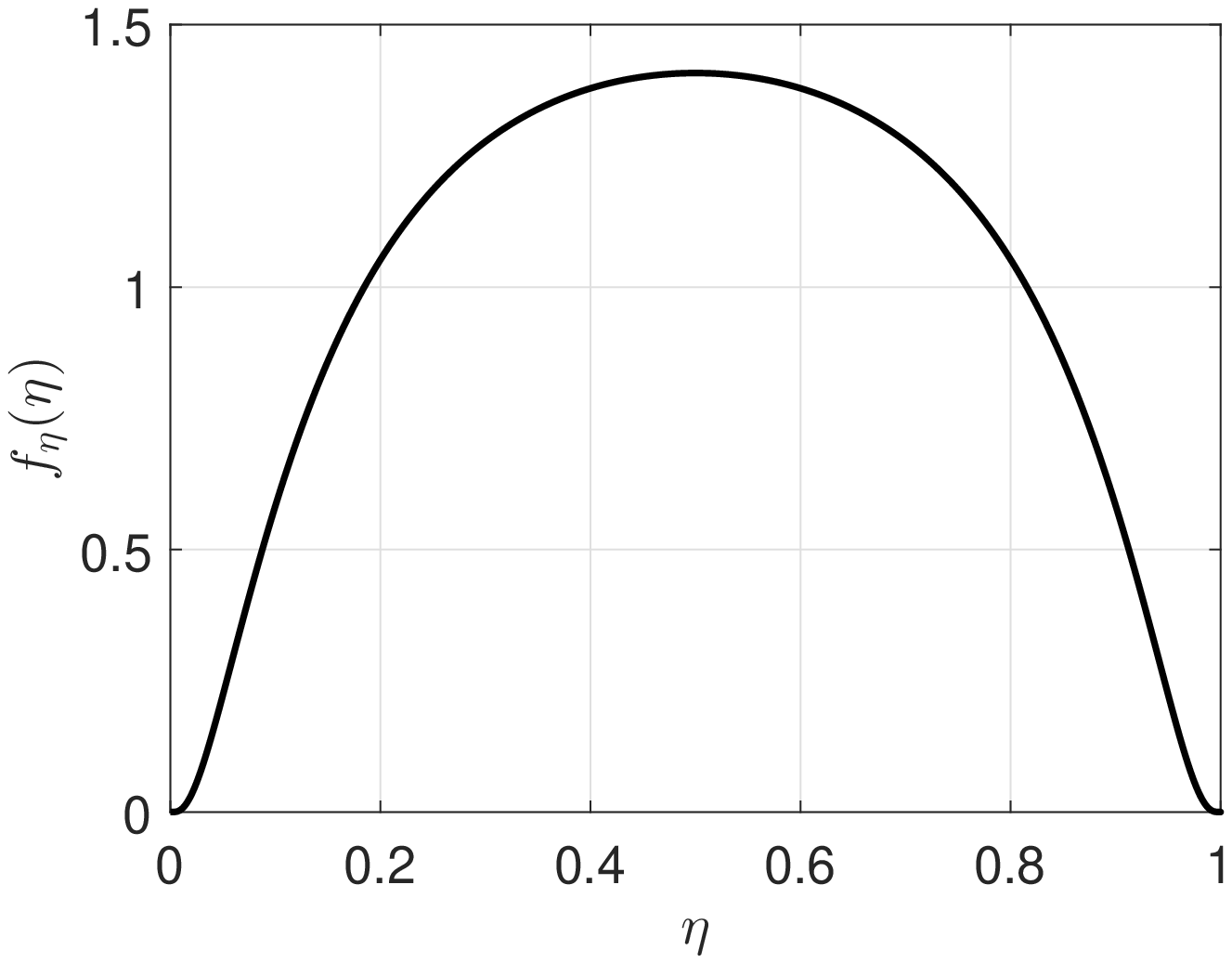}
				\label{fig:feta_plot_sep1}
			}	
			\caption{Illustration of the posterior distribution for two close univariate normal distributions.}
			\label{fig:feta_illustration_plot_sep1}
		\end{center}
	\end{figure*} 
	\begin{figure*}[!t]
		\begin{center}
			\subfloat[Class distribution]{
				\includegraphics[width=0.33\textwidth]{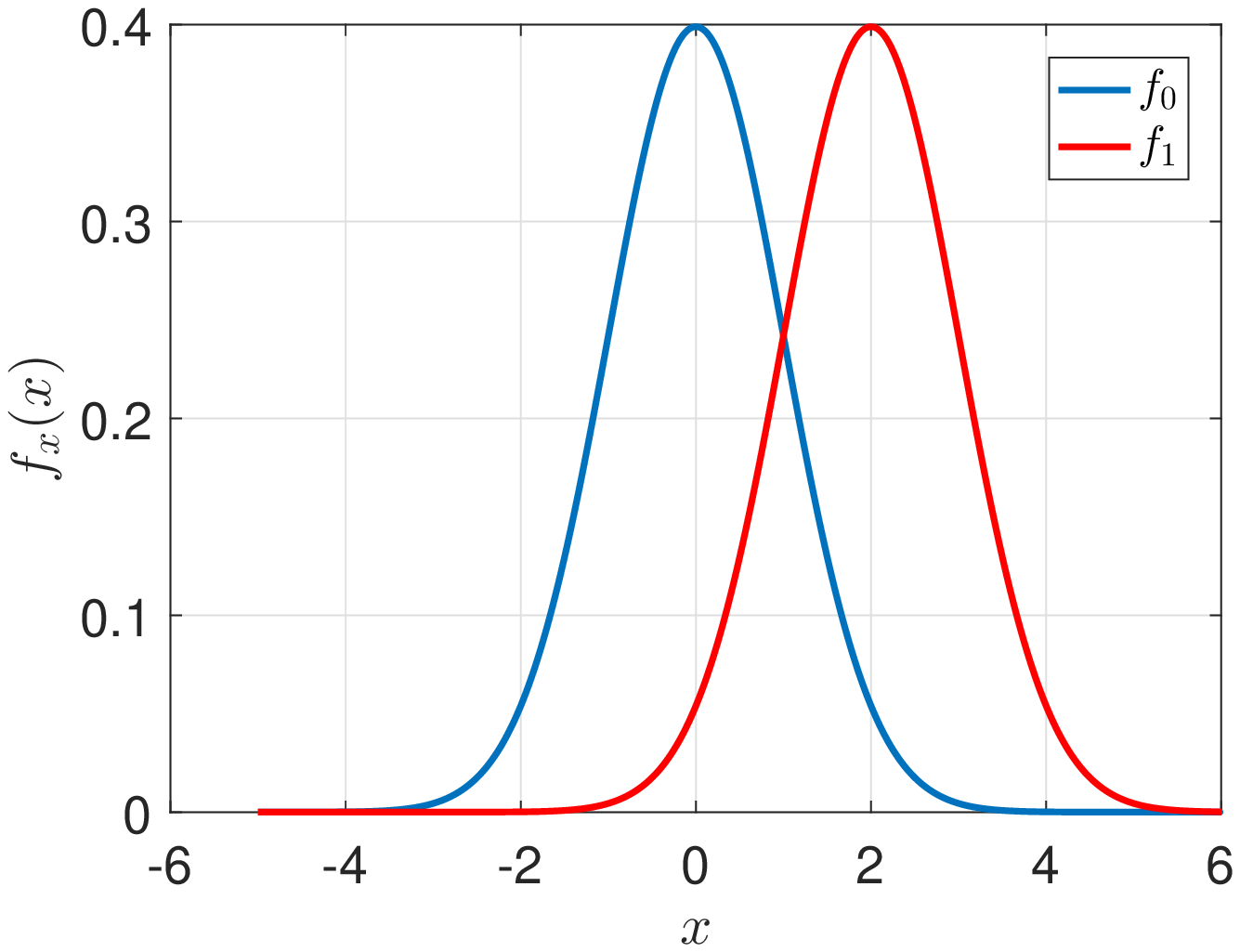}
				\label{fig:f_plot_sep2}
			}	
			\subfloat[Class 1 Posterior]{		
				\includegraphics[width=0.33\textwidth]{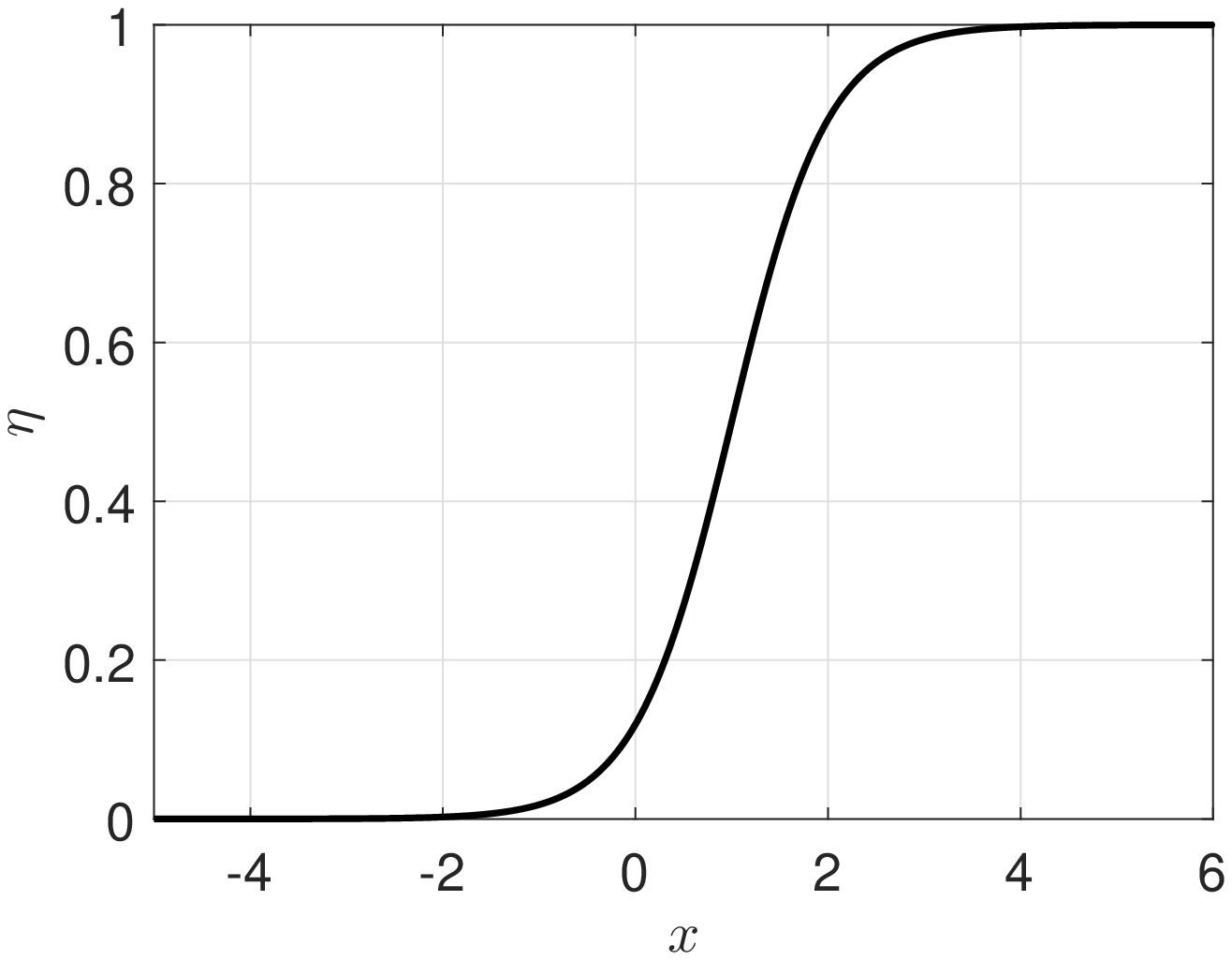}
				\label{fig:eta_plot_sep2}
			}
			\subfloat[Distribution of class 1 posterior likelihoods]{
				\includegraphics[width=0.33\textwidth]{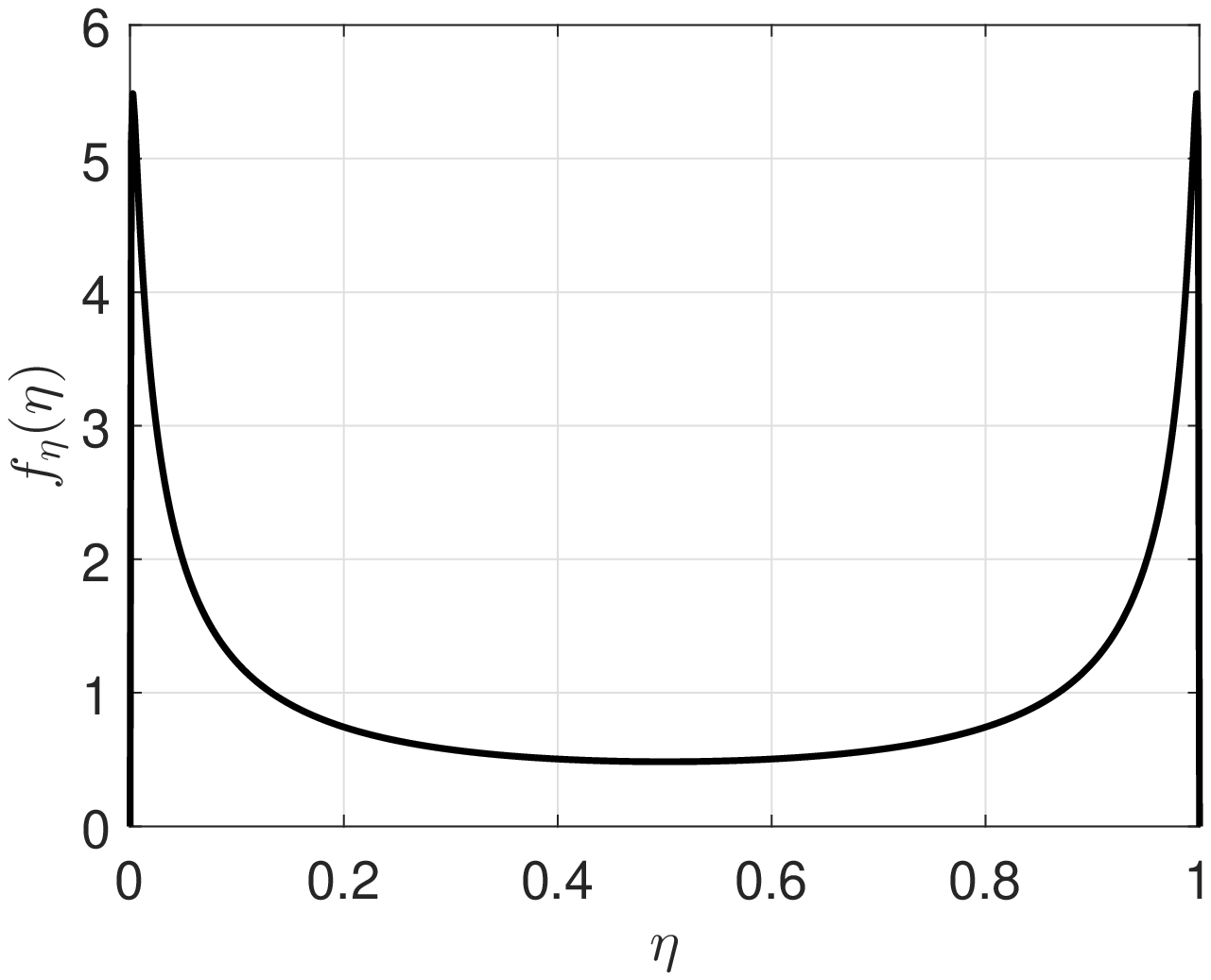}
				\label{fig:feta_plot_sep2}
			}	
			\caption{Illustration of the posterior distribution for two separated univariate normal distributions.}
			\label{fig:feta_illustration_plot_sep2}
		\end{center}
	\end{figure*} 
	We initially introduced $\eta$ as the posterior likelihood function for class 1, which we showed in (\ref{eq:posterior}) can be represented as a function of the underlying distributions. When this function's input is a known point $\mathbf{x}^*$, $\eta(\mathbf{x}^*)$ represents the probability that $y=1$ given that $\mathbf{x}=\mathbf{x}^*$. However, if the input is a random variable $\mathbf{x}$, then $\eta$ is also a random variable, which is distributed according to $f_{\eta}(\eta)$. 
	
	Figure \ref{fig:feta_illustration_plot_sep1} illustrates $f_{\eta}(\eta)$ for two univariate normal distributions $f_0(x)\sim N(0,1)$ and $f_1(x)\sim N(1,1)$. Figure \ref{fig:f_plot_sep1} displays the two class distributions across $x$, Figure \ref{fig:eta_plot_sep1} displays $\eta(x)$ as a function of $x$,  and Figure \ref{fig:feta_plot_sep1} displays $f_{\eta}(\eta)$ as a function of $\eta$. We see from this illustration that while, $\eta(x)$ is close to $0$ or $1$ across most of the region of $x$ that is displayed, $\eta(x)$ remains close to $0.5$ in the regions where $f_x(x)$ is greatest. As a results $f_{\eta}(\eta)$ is roughly bell-shaped, and the likelihood of $\eta$ existing at the extremities (close to $0$ or $1$) is relatively low. Because the probability density in these regions is low, the accuracy of our fit in these regions is less important, and can be given less weight in the fitting routine. Figure \ref{fig:feta_illustration_plot_sep2} repeats this illustration for two well seperated normal distributions. In this case the distribution of $\eta$ is such that $f_{\eta}(\eta)$ is most dense towards the extremities, and therefore they should be given more weight in the fitting routine. Side by side, these two figures present an interesting contrast. Despite the fact that the set of density functions $f_0$ and $f_1$ look quite similar in the two scenarios, the minor difference in separation significantly alters the distribution of $\eta$. 
	
	In practice the underlying distributions $f_0$ and $f_1$ are unknown, and as a result, $f_{\eta}(\eta)$ is unknown as well. However, if we were able to sample $f_{\eta}(\eta)$  at $\eta=\tilde{\eta}_i$, a more direct method of minimizing $e_T$ would be to solve 
	\begin{equation}
	\begin{aligned}
	&w_0,...,w_k=\\ &\underset{w_0,...,w_k}{\mathrm{argmin}}\sum_{i=1}^{\tilde{N}}\Big | g(\tilde{\eta}_i)-\sum_{r=0}^{k}w_r h_{r,k}(\tilde{\eta}_i)\Big |^2 \hat{f}_{\eta}(\tilde{\eta}_i)\Delta_{\tilde{\eta}}
	+\frac{\lambda}{k}\sum_{r=0}^{k}w_r^2,
	\end{aligned}
	\label{eq:fitting_procedure3}
	\end{equation}
	where $\Delta_{\tilde{\eta}} = \tilde{\eta}_{i+1} - \tilde{\eta}_i$.

	Below we show that $\rho_{r,k,N}(\mathbf{X})$,  the statistic previously defined in Theorem 1 can be used to sample the PDF of $\eta$. This result is stated in Theorem 3.
	\begin{mydef} As $N\rightarrow\infty$ and $k\rightarrow \infty$ in a linked manner such that $\frac{k}{N}\rightarrow 0$ and $\frac{r}{k}\rightarrow \eta^*$
		\begin{equation*}
		k\rho_{r,k,N}(\mathbf{X})\rightarrow f_{\eta}(\eta^*). 
		\end{equation*}
		\label{thm:f_eta}
	\end{mydef}

	Theorem \ref{thm:f_eta} is useful as it provides a method of sampling $f_{\eta}(\eta)$ that doesn't depend on estimates of the underlying density functions $f_0$ and $f_1$. Using this result, we can estimate the density of the posterior at point $\tilde{\eta}_i$ as
	\begin{equation}
	\hat{f}_{\eta}(\tilde{\eta}_i)=\tilde{k}_i\rho_{\tilde{r}_i,\tilde{k}_i,N}(\mathbf{X})
	\label{eq:f_eta_hat}
	\end{equation}
	where $\tilde{r}_i$ and $\tilde{k}_i$ are integers selected such that $\tilde{\eta}_i=\frac{\tilde{r}_i}{\tilde{k}_i}$ for $i=1,2,...,\tilde{N}$. Sampling at exactly $\tilde{\eta}_i$ may not always be possible since the maximum value of $k$ is limited by the sample size, and $k$ determines the resolution of the sampling scheme. Even if it is possible, it may not be desirable to recalculate $\rho_{\tilde{r}_i,\tilde{k}_i,N}(\mathbf{X})$ for different values of $k_i$ because of the computational burden. 	
	To overcome these problems we can design our approach such that we utilize the same set of test statistics $\rho_{r,k,N}(\mathbf{X})$ in the estimation of the posterior distribution as are used in the estimation of the basis set. One way to do this is to assign the set of discretized posteriors 
	\begin{equation}
	\boldsymbol{\tilde{\eta}}=[0,\frac{1}{k},\frac{2}{k},...,1]
	\label{eq:tilde_eta}
	\end{equation}
	so that it is straightforward to calculate $\hat{f}_{\eta}(\eta)$ from the known values of  $\rho_{r,k,N}(\mathbf{X})$. An alternate approach is to leave $\tilde{\eta}$ unchanged and interpolate $\hat{f}_{\eta}(\eta)$ to determine its value at the desired points. The advantage of this approach is that it doesn't constrain how $\eta$ is sampled. Throughout the rest of this paper, we will employ the latter method and solve for $\hat{f}_{\eta}(\eta)$ by linearly interpolating between its values on the discretized set \eqref{eq:tilde_eta}.
	
	Because this density-weighted fitting routine more directly minimizes the approximation error of the final estimate, we expect it to generally outperform the uniform method, particularly in cases where the density of the posterior is highly non-uniform and where the desired posterior mapping function $g(\eta)$ is difficult to model using the proposed basis set. This hypothesis will be verified in Sections \ref{sec:divergnece_estimation} and \ref{sec:bound_estimation}, when we empirically evaluate our methods with real data. This improvement comes at a computational cost since the weights are now data-dependent, they must be calculated online, whereas for the uniform method they can be calculated offline and stored. Solving for the $k$ weights can be be implemented with $\mathcal{O}(k^3)$ time complexity \cite{efron2004least}. Current state-of-the-art algorithms for $k$-NN graphs can be implemented with $\mathcal{O}(N\log N)$ time complexity \cite{arya1998optimal}. As a result the complexity for the Convex (uniform) method is $\mathcal{O}(N\log N)$ assuming that the weights are available. The complexity for the Convex (density weighted) method is $\mathcal{O}(k(N)^3 + N\log N)$ since the weights must be determined for each new dataset.

	%
	
	%
	%
	
	%
	%

	\section{Divergence Estimation}
	\label{sec:divergnece_estimation}
	In this section, we show how the proposed method can be applied to estimating three $f$-divergences, the Hellinger distance, the KL-divergence, and the $D_p$-divergence, directly from data. 
	
	\subsection{Hellinger Distance}
	\label{ssec:Hellinger}
	The Hellinger distance squared is an $f$-divergence used to quantify the dissimilarity between two probability distributions and is calculated by 
	\begin{equation}
	H^2(f_0,f_1)=\frac{1}{2}\int\Big (\sqrt{f_0(\mathbf{x})}-\sqrt{f_1(\mathbf{x})}\Big)^2d\mathbf{x}.
	\end{equation}
	Using the approach proposed in Section \ref{sec:fitting_routine}, we can estimate $H^2(f_0,f_1)$ by fitting weights to the posterior mapping function 
	\begin{equation}
	g_H(\eta)=\frac{1}{2}\Bigg(\sqrt{\frac{\eta}{p_1}}-\sqrt{\frac{1-\eta}{p_0}}\Bigg)^2.
	\end{equation} 
	To evaluate the efficacy of this method, we conduct four different experiments in which we attempt to estimate the Hellinger distance between two distributions from finite sample data. In the first three experiments, both distributions are normally distributed according to $f(\mathbf{x})\sim N(\mu \mathbf{1}_d,\mathbf{\Sigma}_d)$, where
	\begin{equation}
	\mathbf{\Sigma}_d=\begin{bmatrix}
	\sigma_{1,1} & \sigma_{1,2}  & \dots & \sigma_{1,d}\\ 
	\sigma_{2,1} & \sigma_{2,2} & \dots & \sigma_{2,d} \\ 
	\vdots & \vdots & \ddots & \vdots \\ 
	\sigma_{d,1} & \sigma_{d,2} & \dots & \sigma_{d,d}
	\end{bmatrix},
	\end{equation}
	for $\sigma_{i,j}=\beta^{|i-j|}$. The first experiment evaluates the most basic case where both Gaussians have spherical covariance. The second experiment considers the case where there exists a strong fixed dependency between adjacent dimensions by using an elliptical covariance structure. The third experiment evaluates the case where this dependency between adjacent dimensions is now dependent on which class the data is drawn from. In the fourth experiment, we return to linearly independent dimensions and consider the case where one of the distributions isn't normally distributed, but instead uniformly distributed within a $d$-dimensional hypercube
	\begin{equation}
	f(\mathbf{x})= \left\{
	\begin{array}{ll}
	\frac{1}{(2\beta)^d} & \mathbf{x} \in [\mu-\beta,\mu+\beta]^d\\
	0 & \mathrm{otherwise}.\\
	\end{array} 
	\right.
	\end{equation}
	A detailed description of the distribution types and parameter setting used in each of the four experiments is presented in Table \ref{tab:experiment_outline}.	In addition to using the proposed method, we also estimate the Hellinger distance using two different plug-in estimators, one based on a parametric density estimator that assumes the data is normally distributed and one based on a $k$-NN density estimate of the underlying distributions. The proposed method is implemented using $\lambda=0.01$ and $k=10$, while setting $\boldsymbol{\tilde{\eta}}=[0,0.01,\dots,1]$. To calculate the $k$-NN estimate, we use the universal divergence estimation approach described in \cite{sutherland2012kernels} and implemented in the ITE toolbox \cite{szabo14information}. This method allows us to fix $k=10$ and still achieve an asymptotically consistent estimate of the divergence. 
	
	In each of the first three experiments, the parametric model shows the highest rate of convergence as expected, although in experiment two it is slightly outperformed at smaller sample sizes by the proposed method. In the fourth experiment, when the assumption of Gaussianity in $f_1$ no longer holds, the parametric solution is significantly biased and as a result, is outperformed by both non-parametric methods at higher sample sizes ($N>2000$). Relative to the $k$-NN plug-in estimator, the proposed method performs slightly worse in experiment 1, slightly better in experiment 2 and significantly better in experiments 3 and 4, with the results being relatively consistent across the various sample sizes. The improvement in performance shown in experiments 3 and 4 suggests that the proposed method offers the greatest benefit when there exists differences in the shapes of the two underlying distributions. Though the density-weighted procedure consistently outperformed the uniform method, the observed improvement was relatively minor in these experiments.
	\begin{table}[]
		\centering
		\caption{Experiment overview table.}
		\label{tab:experiment_outline}
		\begin{tabular}{l|l|l|l|l|l|l|}
			\cline{2-7}
			& \multicolumn{3}{l|}{$f_0(\mathbf{x})$} & \multicolumn{3}{l|}{$f_1(\mathbf{x})$}   \\ \cline{2-7} 
			& Family      & $\mu$      & $\beta$     & Family  & $\mu$                & $\beta$ \\ \hline
			\multicolumn{1}{|l|}{Experiment 1} & Normal      & 0          & 0           & Normal  & $\sqrt{\frac{1}{3}}$ & 0       \\
			\multicolumn{1}{|l|}{Experiment 2} & Normal      & 0          & 0.8         & Normal  & $\sqrt{\frac{1}{3}}$ & 0.8     \\
			\multicolumn{1}{|l|}{Experiment 3} & Normal      & 0          & 0.8         & Normal  & $\sqrt{\frac{1}{3}}$ & 0.9     \\
			\multicolumn{1}{|l|}{Experiment 4} & Normal      & 0          & 0           & Uniform & 0                    & 3       \\ \hline
		\end{tabular}
	\end{table}

	%
	%
	%
	\begin{figure*}[!tb]
		\begin{center}
			\includegraphics[width=0.99\textwidth]{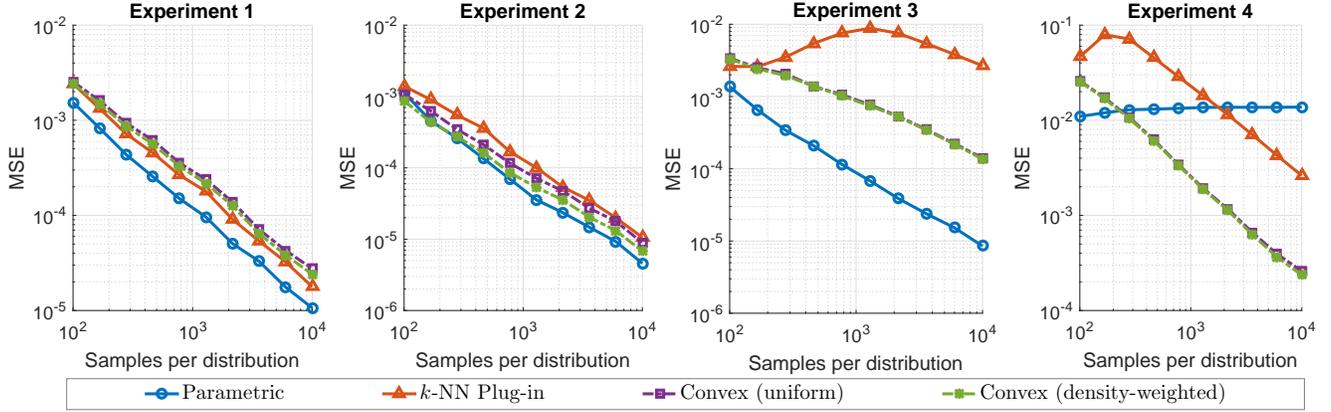}
			\caption{Plots of MSE vs. Sample size in estimating the Hellinger distance for the four different experiments outlined in Table \ref{tab:experiment_outline}.}
			\label{fig:Hellinger_Experiments}
		\end{center}
	\end{figure*}  
	

	\subsection{Kullback Leibler Divergence}
\begin{figure*}[!tb]
		\begin{center}
			\includegraphics[width=0.99\textwidth]{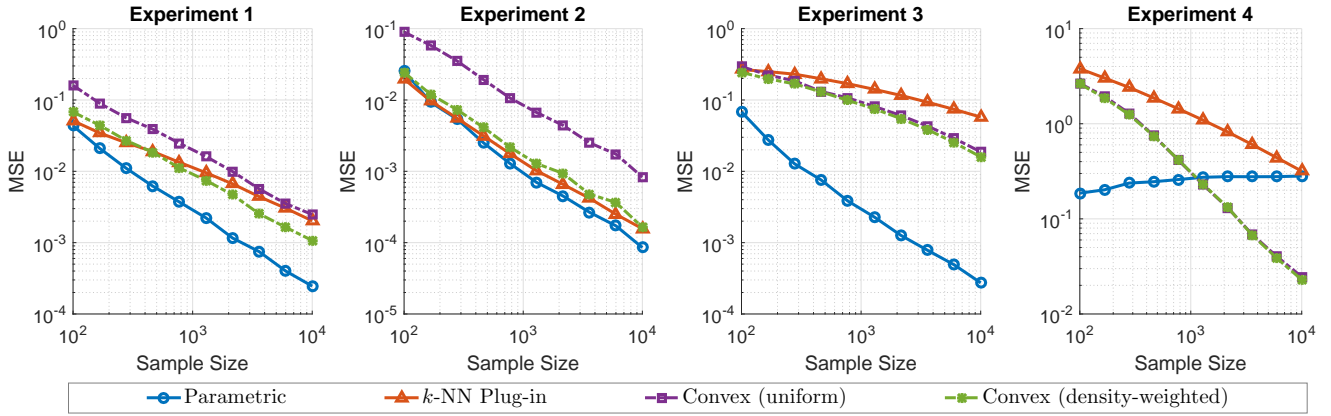}
			\caption{Plots of MSE vs. Sample size in estimating the KL-divergence for the four different experiments outlined in Table \ref{tab:experiment_outline}.}
			\label{fig:KL_Experiments}
		\end{center}
	\end{figure*}

	The Kullback Leibler (KL) divergence \cite{kullback1997information}, also sometimes referred to as the KL risk or relative entropy, is an asymmetric measure of divergence between two probability density functions. Using our regular notation the KL-divergence can be calculated by
	\begin{equation}
	d_{KL}(f_0||f_1)=\int_{-\infty}^{\infty} f_0(\mathbf{x}) \log \bigg( \frac{f_0(\mathbf{x})}{f_1(\mathbf{x})}\bigg)d\mathbf{x}.
	\end{equation}
	While the KL-divergence has the same general purpose as the Hellinger distance, that is to measure the dissimilarity between two probability density functions, it also has several key difference. Firstly since the KL-divergence is asymmetric it doesn't technically qualify as a distance function and $d_{KL}(f_0||f_1)$ isn't necessarily equal to $d_{KL}(f_1||f_0)$. This also means that $d_{KL}(f_0||f_1)$ and $d_{KL}(f_1||f_0)$ will have different posterior mapping functions. We can define the posterior mapping function for $d_{KL}(f_0||f_1)$ as
	\begin{equation}
	g_{KL}^0(\eta)=\frac{1-\eta}{p_0}\log\bigg(\frac{p_1(1-\eta)}{p_0\eta}\bigg)
	\end{equation}
	and the posterior mapping function for $d_{KL}(f_1||f_0)$ as
	\begin{equation}
	g_{KL}^1(\eta)=\frac{\eta}{p_1}\log\bigg(\frac{p_0\eta}{p_1(1-\eta)}\bigg)
	\end{equation}
	such that
	\begin{equation}
	d_{KL}(f_i||f_{|i-1|})=\int_{-\infty}^{\infty}g_{KL}^i(\eta(\mathbf{x}))(p_0f_0(\mathbf{x})+p_1f_1(\mathbf{x}))d\mathbf{x}.
	\end{equation}
	It is worth noting that when $p_0=p_1=0.5$
	\begin{equation}
	g_{KL}^1(\eta)=g_{KL}^0(1-\eta)
	\end{equation}  
	thus $g_{KL}^1(\eta)$ is a reflection of $g_{KL}^0(\eta)$. One challenge presented in modeling the KL-divergence is that the posterior mapping functions are difficult to model due to discontinuities at the end points, since $g_{KL}^0(0)=\infty$ and $g_{KL}^0(1)$ is undefined though 
	\begin{equation}
	\lim\limits_{\eta\rightarrow 1^-}g_{KL}^0(\eta)=0.
	\end{equation}
	Due to their symmetry $g_{KL}^1$ has the same problem at the opposite endpoints. To handle this we simply select our discretized set of posteriors  $\tilde{\eta}_1,\tilde{\eta}_2,...,\tilde{\eta}_{\tilde{N}}$ such that $0 <\tilde{\eta}_1 < \tilde{\eta}_2 <...<\eta_{\tilde{N}} < 1$. For the experiments in this Section, we set \begin{equation}
	\tilde{\eta}_1,\tilde{\eta}_2,\tilde{\eta}_3,...,\tilde{\eta}_{100},\tilde{\eta}_{101}=\epsilon,0.01,0.02,...,0.99,1-\epsilon
	\end{equation} 
	where $\epsilon=10^{-4}$. Using this modified set of discretized posteriors, we repeat the set of four experiments described in Section \ref{ssec:Hellinger} to evaluate the proposed methods ability to estimate the KL-divergence. The results of this experiment are displayed in Figure \ref{fig:KL_Experiments}. Like the estimates of the Hellinger distance, the parametric method generally yielded the best performance in the first three experiments, but suffered from a large asymptotic bias in experiment 4. The proposed method once again significantly outperformed the $k$-NN plug-in estimator in experiments 3 and 4, however the results in experiments 1 and 2 are slightly more nuanced due to the significant difference in performance between the two optimization criteria in these trials. In both of these trials the density-weighted criteria significantly outperforms the uniform method at all sample sizes. In experiment 1 the $k$-NN plug-in estimator outperforms the regular plug-in estimator at all sample sizes, but is outperformed by the density-weighted method for $N>300$. In experiment 2 the $k$-NN plug-in estimator consistently outperforms both proposed methods, however the improvement over the density-weighted method is marginal. 
	%
	%
	
	\subsection{$D_p$-Divergence}
\begin{figure*}[!tb]
		\begin{center}
			\includegraphics[width=0.99\textwidth]{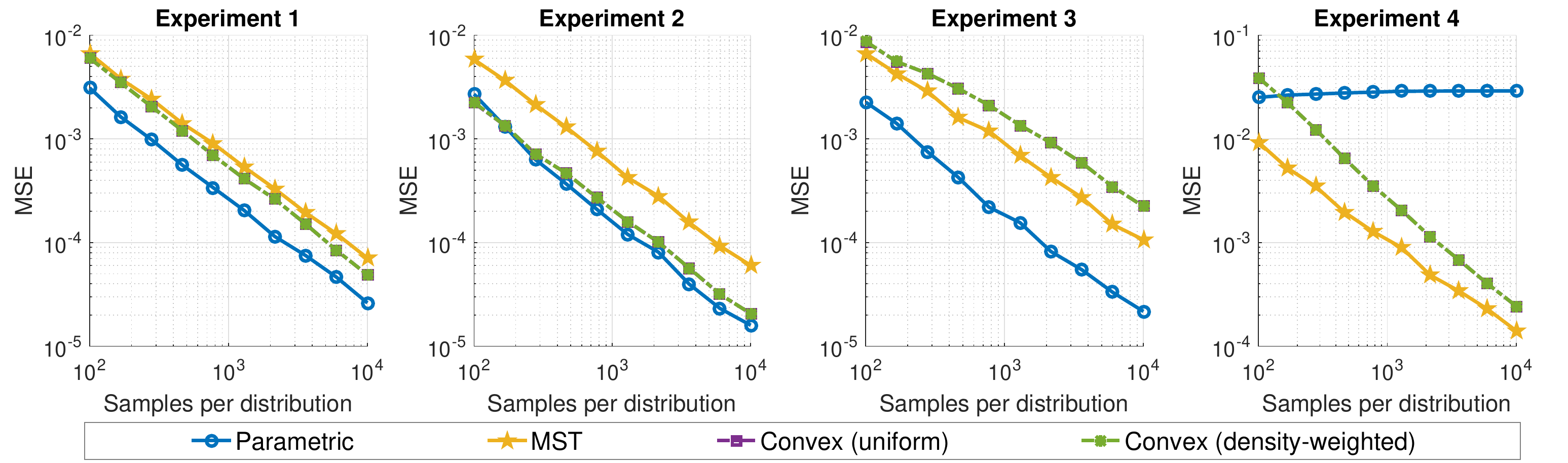}
			\caption{Plots of MSE vs. Sample size in estimating the $D_p$-divergence for the four different experiments outlined in Table \ref{tab:experiment_outline}.}
			\label{fig:Dp_Experiments}
		\end{center}
	\end{figure*}

	The $D_p$-divergence is an $f$-divergence defined by
	\begin{equation}
	\label{eqn:div_HP}
	D_{p_0}(f_0,f_1) = \frac{1}{4p_0p_1}\left [ \int \frac{(p_0 f_0(\mathbf{x})- p_1f_1(\mathbf{x}))^2}{p_0 f_0(\mathbf{x}) + p_1 f_1(\mathbf{x})} d\mathbf{x} - (p_0-p_1)^2 \right].
	\end{equation}
	The $D_p$-divergence has the unique property of being directly estimable from data using minimum spanning trees \cite{berisha2014empirically}. Because of this property, it has been used to form non-parametric estimates of the Fisher information \cite{berisha2014empirical} as well as upper bounds on the Bayes error rate in a range of classification problems \cite{berisha2014empirically,wisler2016multiclass}. The posterior mapping function for the $D_p$-divergence can be defined as
	
	\begin{equation}
	g_{D_p}(\eta)=\frac{(2\eta-1)^2-(2p_0-1)^2}{4p_0(1-p_0)}
	\label{eq:Dp_PMfunction}
	\end{equation}
	which simplifies to $(2\eta-1)^2$ when $p_0=p_1=0.5$. We once again repeat the experiments described in Section \ref{ssec:Hellinger} to evaluate the proposed methods ability to estimate the $D_p$-divergence. This experiment provides the unique opportunity to compare the proposed method to a more traditional direct estimation procedure. The results of this experiment are displayed in Figure \ref{fig:Dp_Experiments}.
	
	As in the previous experiments, the parametric estimate generally performed the best in the first three experiments, but suffered from a large asymptotic bias in experiment 4. The proposed methods perform better than the MST-based estimation in experiments 1 and 2 but worse in experiments 3 and 4. The relative performance of each method in this experiment was largely consistent across the various sample sizes, though the proposed method seems to be converging slightly faster than the MST method in experiment 4 and could possibly exceed its performance given enough samples. Unlike in the previous experiments, we found no difference in performance between the uniform optimization criteria and the density-weighted criteria in this experiment. This is due to the fact that $g_{D_p}(\eta)$ is a polynomial and can be perfectly represented by the proposed basis set, even when $k$ is truncated. Since we are able to achieve a solution where $\epsilon(\eta)=0$ $ \forall \eta$, the density weighting has no impact on the final results.

	\section{Fitting Bounds on Performance}
	\label{sec:bound_estimation}
	The optimization criteria in the proposed fitting routines gives us the ability to not only approximate information-theoretic functions, but to bound them as well. This is especially useful for forming bounds on the Bayes Error Rate (BER). The Bayes error rate represents the optimal classification performance that is achievable for a given pair of class distributions $f_0(\mathbf{x})$ and  $f_1(\mathbf{x})$ with prior probabilities $p_0$ and $p_1$ respectively and can be calculated by 
	\begin{equation}
	\epsilon^{\mathrm{Bayes}} = \! \! \! \! \! \! \! \! \! \! \! \! \! \int\limits_{p_0 f_0(\mathbf{x}) \leq p_1 f_1(\mathbf{x})} \! \! \! \! \! \! \! \! \! \! \! \! p_0 f_0(\mathbf{x}) d\mathbf{x} \ \ + \! \! \! \! \! \! \! \!   \int\limits_{p_1 f_1(\mathbf{x}) \leq p_0 f_0(\mathbf{x})} \! \! \! \! \! \! \! \! \! \! \! \! p_1 f_1(\mathbf{x}) d\mathbf{x}.
	\label{eq:BER}
	\end{equation}
	In essence the BER measures the intrinsic difficulty of a particular classification problem based on the data. A thorough understanding of the BER of a particular problem can help design optimal classifiers. Because of the challenges associated with estimating the BER, much of the literature has focused on generating bounds on the BER \cite{kailath1967divergence,berisha2014empirically}, which are generally formulated in terms of some measure of divergence between the two class distributions. One such bound, the well-known Bhattacharya bound, is given by \cite{kailath1967divergence}
	\begin{equation} \label{eqn:bcbound}
	\frac{1}{2}-\frac{1}{2}\sqrt{1-BC^2(f_0,f_1)} \leq \epsilon^{\mathrm{Bayes}} \leq \frac{1}{2} BC(f_0,f_1),
	\end{equation}
	where 
	\begin{equation}
	BC(f_0,f_1)=1-H^2(f_0,f_1).
	\end{equation}
	While the Hellinger distance here can be estimated via any of the methods discussed in Section \ref{ssec:Hellinger}, parametric estimates are most common. Alternatively \cite{berisha2014empirically} introduced the bounds 
	\begin{equation}
	\frac{1}{2} - \frac{1}{2}\sqrt{D_{\frac{1}{2}}(f_0,f_1)} 
	\leq \epsilon^{\mathrm{Bayes}} \leq   \frac{1}{2} -  \frac{1}{2}D_{\frac{1}{2}}(f_0,f_1) 
	\end{equation}
	where
	\begin{equation}
	D_{\frac{1}{2}}(f_0,f_1) =1-2\int \frac{ f_0(\mathbf{x}) f_1(\mathbf{x})}{f_0(\mathbf{x}) + f_1(\mathbf{x})} d\mathbf{x}.
	\end{equation}
	These bounds have the advantage of being provably tighter than the Bhattacharyya bounds \cite{berisha2014empirically}. Furthermore since $D_{\frac{1}{2}}$ represents a particular case of the $D_p$-divergence, which is estimable directly from data, these bounds bypass the need for density estimation much like the approaches proposed in this paper.  While these bounds are significantly tighter than the Bhattacharyya bounds, they still leave room for improvement. Avi-Itzhak proposed arbitrarily tight bounds on the BER in \cite{avi1996arbitrarily}, however these bounds require density estimation to be employed in practical problems. In this section, we will use a modified version of the previously described fitting routine in order to investigate how tightly we are able to bound the BER using a linear combination of directly estimable basis functions.
	\begin{figure}[hptb]
		\includegraphics[width=0.5\textwidth]{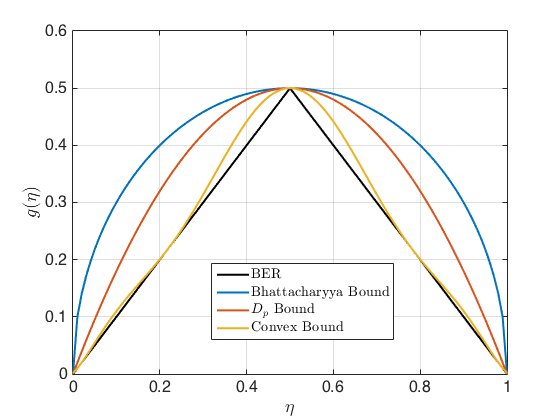}
		\caption{The Bayes error rate along with the three considered upper bounds displayed as a function of $\eta$.}
		\label{fig:bounds_theoryplot}
	\end{figure}
	%
	%
	
	\begin{figure*}[!tb]
		\begin{center}
			\includegraphics[width=0.99\textwidth]{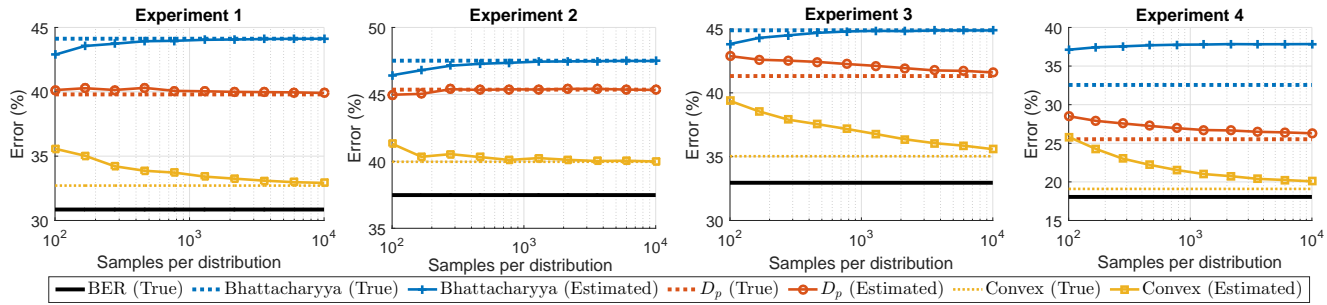}
			\caption{Plots of theoretical and estimate upper bounds on the BER as a function of sample size for the four different experiments outlined in Table \ref{tab:experiment_outline}.}
			\label{fig:UB_Experiments}
		\end{center}
	\end{figure*} 
	Using the fitting routine described in (\ref{eq:fitting_procedure_wRegularization}) to bound the BER, requires that we define $g(\eta)$ appropriately for estimation of the BER, and constrain our fit such that 
	\begin{equation}
	\sum_{r=0}^{k}w_r h_{r,k}(\tilde{\eta}_i) \geq g(\tilde{\eta}_i)  \quad \forall \tilde{\eta}_i .
	\end{equation}
	We can express (\ref{eq:BER}) as
	\begin{equation}
	\begin{aligned}
	\epsilon^{\mathrm{Bayes}}&=\int \min \big[ p_0 f_0(\mathbf{x}),p_1 f_1(\mathbf{x})\big]d\mathbf{x} \\
	&=\int\min \big[ 1-\eta(\mathbf{x}),\eta(\mathbf{x})\big]f_{\mathbf{x}}(\mathbf{x})d\mathbf{x}
	\end{aligned}
	\end{equation}
	so $g(\eta)=\min \big[ 1-\eta,\eta\big]$. Incorporating these changes within the regularized fitting routine described in (\ref{eq:fitting_procedure_wRegularization}) yields
	\begin{equation}
	\begin{aligned}
	&w_0,...,w_k=\\ &\underset{w_0,...,w_k}{\mathrm{argmin}}\frac{1}{\tilde{N}}\sum_{i=1}^{\tilde{N}}\Big | g(\tilde{\eta}_i)-\sum_{r=0}^{k}w_r h_{r,k}(\tilde{\eta}_i)\Big |^2
	+\frac{\lambda}{k}\sum_{r=0}^{k}w_r^2 \\
	&  \text{subject to } \quad \quad \sum_{r=0}^{k}w_r h_{r,k}(\tilde{\eta}_i) \geq g(\tilde{\eta}_i)  \quad \quad \forall \tilde{\eta}_i.
	\end{aligned}
	\label{eq:fitting_procedure_bounds_wRegularization}
	\end{equation}
	
	Figure \ref{fig:bounds_theoryplot} compares the theoretical values of each of these bounds as a function of $\eta$. These results indicate that the proposed method offers much tighter theoretical bounds than the other two methods, however this bound is based on the asymptotic properties of the proposed basis set and doesn't consider the limitations of a finite sample estimate. 
	
	We evaluate the finite-sample performance of this method by calculating each of the described bounds across the four experiments described in Table \ref{tab:experiment_outline}. Figure \ref{fig:UB_Experiments} displays the ground truth value of the BER, along with the theoretical and estimated values for each of the three bounds (Bhattacharyya, $D_p$, and convex) across sample sizes ranging from 100 to 10000. The Bhattacharyya bound is calculated based on a parametric plug-in estimator which assumes both class distributions to be normally distributed. The $D_p$ bound is calculated from the Friedman-Rafsky test statistic using the approach described in \cite{berisha2014empirically}. Finally the convex bound is calculated as a linear combination of the proposed directly estimable basis functions using weights optimized according to (\ref{eq:fitting_procedure_bounds_wRegularization}). The results of this experiment are largely consistent across the four experiments, the convex method yields the tightest bound, followed by the $D_p$ bound, and finally the Bhattacharyya bound. Except for the Bhattacharyya bound in experiment 4, which is estimated parametrically, all of the bounds appear to converge to their asymptotic solution. While the convex bound generally offers a slightly slower convergence rate than the other two solutions, it remains tighter than the other two bounds across all sample sizes.
	
	In order to further validate this bound we repeat one of the experiments conducted in \cite{berisha2014empirically} by evaluating the proposed bound along with the Mahalanobis bound, the Bhattacharyya bound, and the $D_p$ bound on two 8-dimensional Gaussian data sets described in \cite{fukunaga1990introduction}. The mean and standard deviations of $f_0$ and $f_1$ for the two data sets are described in Table \ref{tab:BERDataSets}, and all dimensions are independent. These data sets allow us to analyze the tightness and validity of the bounds in a higher dimensional setting. For this experiment the sample size was fixed at $N=1000$ and only the empirical value of each of the bounds was evaluated. Table \ref{table:Fukunage_Results} displays the mean and standard deviation of each bound calculated across 500 Monte Carlo iterations for each of the two data sets. In both data sets the convex method provides the tightest bounds on the BER.
	
	\begin{table}
		\caption{Parameters for 2 8-dimensional Gaussian data sets for which the Bayes error rate is known (from \cite{fukunaga1990introduction}).}
		\resizebox{\columnwidth}{!}{
			\begin{tabular}{c  c | c  c  c  c  c  c  c  c  }
				\hline
				\multirow{4}{*}{$\mathcal{D}_0$} & $\boldsymbol{\mu_0}$ & 0& 0& 0& 0& 0& 0& 0& 0 \\
				& $\boldsymbol{\sigma_0}$ & 1& 1& 1& 1& 1& 1& 1& 1 \\
				& $\boldsymbol{\mu_1}$ & 2.56 & 0& 0& 0& 0& 0& 0& 0 \\
				& $\boldsymbol{\sigma_1}$ & 1& 1& 1& 1& 1& 1& 1& 1 \\ \hline
				\multirow{4}{*}{$\mathcal{D}_2$} & $\boldsymbol{\mu_0}$ & 0& 0& 0& 0& 0& 0& 0& 0 \\
				& $\boldsymbol{\sigma_0}$ & 1& 1& 1& 1& 1& 1& 1& 1 \\
				& $\boldsymbol{\mu_1}$ & 3.86 & 3.10 & 0.84 & 0.84 & 1.64 & 1.08 & 0.26& 0.01 \\
				& $\boldsymbol{\sigma_1}$ & 8.41 & 12.06& 0.12 & 0.22 & 1.49 & 1.77& 0.35 & 2.73 \\ \hline
			\end{tabular}
		}
		\label{tab:BERDataSets}
	\end{table}

	\begin{table}
		\caption{Comparing upper bounds on the Bayes error rate for the multivariate Gaussians defined in Table \ref{tab:BERDataSets}.}
		\centering
		\resizebox{\columnwidth}{!}{
			\begin{tabular}{l c c }
				\hline\hline
				& Data 1 & Data 2 \\ [0.5ex] 
				\hline
				Actual Bayes Error & 10\% & 1.90\% \\
				Mahalanobis   Bound & 18.90\% $\pm$ 0.55\% & 14.07 \% $\pm$ 0.45 \% \\
				Bhattacharyya  Bound & 21.74\% $\pm$ 0.87 \% & 4.68 \% $\pm$ 0.27 \% \\
				$D_p$ Bound & 16.51 \% $\pm$ 1.07\% & 3.99 \% $\pm$ 0.52 \% \\
				Convex Bound & \bf{14.17 \%} $\pm$ \bf{0.86\%} &  \bf{3.87\%} $\pm$ \bf{0.43\%}  \\
				\hline
			\end{tabular}
		}
		\label{table:Fukunage_Results}
	\end{table}
	
	\section{Conclusion}
	\label{sec:conclusion}
	This paper introduces a novel method for estimating density functionals which utilizes a set of directly estimable basis functions. The most appealing feature of the proposed method is its flexibility. Where previous methods of direct estimation are generally only applicable to a specific quantity, we show that the basis set can be used to  generate an asymptotically consistent estimate of a broad class of density functionals, including all $f$-divergences and the Bayes error rate. We validate these findings by experimentally evaluating the proposed method's ability to estimate three different divergences (the KL-divergence, the Hellinger distance, and the $D_p$-divergence) for four pairs of multivariate probability density functions. The results reveal that the proposed method performs competitively with other non-parametric divergence estimation methods, and seems to outperform them in cases where the data from the two distributions have different covariance structures or belong to different families. Additionally we demonstrate how the method can be modified to generate empirically-estimable bounds on the Bayes error rate that are much tighter than existing bounds.
	
	Future work could focus on studying the finite-sample properties of the basis set proposed in this paper, since this represents a major source of error for the proposed methodology. An improved understanding of these properties could enable us to refine the regularization term in our optimization criteria to more accurately model each weights contribution to the estimation error or to develop ensemble methods, like those in \cite{moon2014multivariate}, for estimating the individual basis functions. Another worthwhile future direction would be on determining an orthogonal version of the Bernstein basis expansion since this would simplify solving for the weights of the expansion.
	
	\appendices
	\section{Proof of Theorem \ref{thm:Basis1} }
	\label{sec:proof_thm1}
	Aspects of this proof mirror the methods used to prove the asymptotic convergence of the $k$-NN error rate in pages 62-70 of \cite{devroye2013probabilistic}, please consult this text for additional details. Construction of this proof requires the introduction of an auxilary variable $\mathbf{u}$. We define the augmented data set $[\mathbf{X},\mathbf{y},\mathbf{u}]$, where $\mathbf{X}$ is as defined previously, $\mathbf{u}$ is a set of i.i.d. random variables which are uniformly distributed in $[0,1]$, and $\mathbf{y}$ is defined such that
	\begin{equation}
	y_i= \left\{
	\begin{array}{ll}
	1 & u_i\leq \eta(\mathbf{x}_i) \\
	0 & \mathrm{otherwise}.\\
	\end{array} 
	\right.
	\end{equation}  
	Now, let us consider an arbitrary point $\mathbf{x}^*$ in the support of $f_{\mathbf{x}}(\mathbf{x})$. For each instance $\mathbf{x}_i$, we can define an alternate label 
	\begin{equation}
	y_i'= \left\{
	\begin{array}{ll}
	1 & u_i\leq \eta(\mathbf{x}^*) \\
	0 & \mathrm{otherwise}.\\
	\end{array} 
	\right.
	\end{equation}
	Unlike the real labels $y_i$, these alternate labels $y'_i$ have no dependency on $\mathbf{x}_i$; they depend only on the fixed point $\mathbf{x}^*$. From these alternate labels we construct the alternate statistic 
	\begin{equation}
	\Phi_k'(\mathbf{x}^*)=\sum_{i: \mathbf{x}_i\in\mathcal{N}_k(\mathbf{x}^*)}^{}y_i'.
	\end{equation}
	Since $\Phi_k'(\mathbf{x}^*)$ is simply the sum of $k$ i.i.d. Bernoulli random variables, we can express the probability that $\Phi_k'(\mathbf{x}^*)=r$ as 	
	\begin{equation} 
	P[\Phi_{k}'(\mathbf{x})=r | \mathbf{x} = \mathbf{x}^*]=\dbinom{k}{r}\eta^r(\mathbf{x}^*)(1-\eta(\mathbf{x}^*))^{k-r}.
	\label{eq:bino1}
	\end{equation}
	We can upper bound the likelihood that $\Phi'_{k}(\mathbf{x}^*)\neq\Phi_{k}(\mathbf{x}^*)$ by
	\begin{equation}
	P\left[\Phi'_{k}(\mathbf{x}^*)\neq\Phi_{k}(\mathbf{x}^*)\right]\leq \sum_{i: \mathbf{x}_i\in\mathcal{N}_k(\mathbf{x}^*)}^{}P[y_i\neq y_i'].
	\label{eq:Phi_Bound}
	\end{equation}
	Using the definitions of $y_i$ and $y_i'$ above, \eqref{eq:Phi_Bound} can be expressed in terms of the difference between the posterior likelihood at $\mathbf{x}^*$ vs. $\mathbf{x}_i$ as
		\begin{equation}
		P\left[\Phi'_{k}(\mathbf{x}^*)\neq\Phi_{k}(\mathbf{x}^*)\right]\leq \sum_{x_i\in\mathcal{N}_k(\mathbf{x}^*)}^{}E\big[|\eta(\mathbf{x}^*)-\eta(\mathbf{x}_i)|\big].
		\label{eq:lemma_5p2}
		\end{equation}  
		 Using Lemma 5.4 in \cite{devroye2013probabilistic}, we can show that
	\begin{equation}
	\sum_{x_i\in\mathcal{N}_k(\mathbf{x}^*)}^{}E\big[|\eta(\mathbf{x}^*)-\eta(\mathbf{x}_i)|\big]\rightarrow 0
	\label{eq:lemma_5p4}
	\end{equation}
	as $N\rightarrow \infty$ whenever $\frac{k}{N}\rightarrow 0$. Combining \eqref{eq:lemma_5p2} and \eqref{eq:lemma_5p4}
	\begin{equation}
	P\left[\Phi'_{k}(\mathbf{x}^*)\neq\Phi_{k}(\mathbf{x}^*)\right]\rightarrow 0
	\label{eq:Equivalence}
	\end{equation}  
	as $N\rightarrow \infty$ whenever $\frac{k}{N}\rightarrow 0$. Furthermore, since convergence in probability implies convergence in distribution \cite{van2000asymptotic}, 
	\begin{equation}
	\lim\limits_{N\rightarrow \infty }P[\Phi_k(\mathbf{x}^*)=r]= P[\Phi'_{k}(\mathbf{x}^*)=r].
	\label{eq:conv_dist}
	\end{equation}
	 Using \eqref{eq:conv_dist}, we can simplify the expectation of $I_{r,k}(\mathbf{x})$ to 
	 \begin{equation}
	 \begin{aligned}
	 \lim\limits_{N\rightarrow \infty }E[I_{r,k}(\mathbf{x})]&=\lim\limits_{N\rightarrow \infty }E[E[I_{r,k}(\mathbf{x})|\mathbf{x}=\mathbf{x}']]\\
	 &=E[P[\Phi_k'(\mathbf{x})=r|\mathbf{x}=\mathbf{x}']]\\
	 =\int  \dbinom{k}{r}&\eta(\mathbf{x}')^r(1-\eta(\mathbf{x}'))^{k-r}f_{\mathbf{x}}(\mathbf{x}')d\mathbf{x}'=\rho^*
	 \end{aligned}
	 \end{equation}
	 where $\mathbf{x}'$ represents a random variable independent of all instances $\mathbf{x}_i$ and distributed according to $f_{\mathbf{x}}(\mathbf{x}')$.	 Now, let us evaluate the expression 
	\begin{equation}
	\begin{aligned}
	E\big[(\rho_{r,k,N}(\mathbf{X})-\rho^*)^2\big]&=E[(\rho_{r,k,N}(\mathbf{X}))^2]\\-2&E[\rho^*\rho_{r,k,N}(\mathbf{X})]+E[(\rho^*)^2].
	\label{eq:MSE_expansion}
\end{aligned}
\end{equation}
Beginning with the first term in \eqref{eq:MSE_expansion}
\begin{equation}
\begin{aligned}
E[(\rho_{r,k,N}(\mathbf{X}))^2]=E\bigg[\frac{1}{N^2}\sum_{\mathbf{x}_i\in\mathbf{X}}^{}\sum_{\mathbf{x}_j\in\mathbf{X}}^{}I_{r,k}(\mathbf{x}_i)I_{r,k}(\mathbf{x}_j)\bigg]\\
=\frac{1}{N^2}E\bigg[\sum_{\mathbf{x}_i\in\mathbf{X}}^{}I_{r,k}(\mathbf{x}_i)+\frac{1}{N^2}\mathop{\sum\sum}_{\mathbf{x}_i,\mathbf{x}_j\in\mathbf{X}; i\neq j}I_{r,k}(\mathbf{x}_i)I_{r,k}(\mathbf{x}_j)\bigg]\\
=\frac{1}{N}E\big[I_{r,k}(\mathbf{x}_i)\big]+\frac{N^2-N}{N^2}E\big[I_{r,k}(\mathbf{x}_i)I_{r,k}(\mathbf{x}_j)\big]
\label{eq:term_1_1}
\end{aligned}
\end{equation}
Note that the random variables $I_{r,k}(\mathbf{x}_i)$ and $I_{r,k}(\mathbf{x}_j)$ are conditionally independent if $B$ is true, where $B=\{\mathbf{x}_i \in \mathbf{X} \setminus \mathcal{N}_k(\mathbf{x}_j) \cap \mathbf{x}_j \in \mathbf{X} \setminus \mathcal{N}_k(\mathbf{x}_i)\}$. Since every instance in $\mathbf{X}\setminus \mathbf{x}_i$ is equally likely to be in $\mathcal{N}_k(\mathbf{x}_i)$, we can bound the probability of the complement of $B$ by $P[\bar{B}]\leq 2k/N$. From this we know that as $N\rightarrow \infty$, $P[B]\rightarrow 1$ and as a result
\begin{equation}
\begin{aligned}
\lim\limits_{N\rightarrow \infty}E\big[I_{r,k}(\mathbf{x}_i)I_{r,k}(\mathbf{x}_j)\big]&=\lim\limits_{N\rightarrow \infty}E\big[I_{r,k}(\mathbf{x}_i)\big]E\big[I_{r,k}(\mathbf{x}_j)\big]\\
&=(\rho^*)^2.
\end{aligned}
\end{equation}
Plugging this into \eqref{eq:term_1_1} yields
\begin{equation}
\lim\limits_{N\rightarrow \infty}E[(\rho_{r,k,N}(\mathbf{X}))^2]=\lim\limits_{N\rightarrow \infty}\frac{1}{N}\rho^*+\frac{N-1}{N}(\rho^*)^2.
\end{equation}
Similarly the second term simplifies to
\begin{equation}
\begin{aligned}
\lim\limits_{N\rightarrow \infty}E[\rho_{r,k,N}(\mathbf{X})\rho^*]=\lim\limits_{N\rightarrow \infty}\rho^*E\bigg[\frac{1}{N}\sum_{\mathbf{x}_i\in\mathbf{X}}^{}I_{r,k}(\mathbf{x}_i)\bigg]=(\rho^*)^2
\end{aligned}
\end{equation}
and the third term clearly equals $(\rho^*)^2$. Substituting these results into \eqref{eq:MSE_expansion} yields
\begin{equation}
\begin{aligned}
\lim\limits_{N\rightarrow \infty}E\big[(\rho_{r,k,N}(\mathbf{X})-\rho^*)^2\big]=&\lim\limits_{N\rightarrow \infty}\frac{1}{N}\rho^*
+\frac{N-1}{N}(\rho^*)^2\\
&-2 (\rho^*)^2+(\rho^*)^2\\
=&\lim\limits_{N\rightarrow \infty}\frac{\rho^*-(\rho^*)^2}{N}= 0.
\label{eq:MSE_expansion2}
\end{aligned}
\end{equation}
%
%
%
%
%
	\section{Proof of Theorem \ref{thm:final}}
	
	Starting with the expression
	\begin{equation}
	\lim\limits_{k\rightarrow \infty}\lim_{\substack{N\to \infty\\ k/N\to 0}} \hat{G}_{k,N}(\mathbf{X})=
	\lim\limits_{k\rightarrow \infty}\lim_{\substack{N\to \infty\\ k/N\to 0}} \sum_{r=0}^{k}g\Big(\frac{r}{k}\Big)\rho_{r,k,N}(\mathbf{X})
	\label{eq:prf_3_1}
	\end{equation}
	we first evaluate the limit with respect to $N$. Since these conditions mirror those of Theorem \ref{thm:Basis1} and $g(r/k)$ is independent of $N$, we can rewrite \eqref{eq:prf_3_1} as
	\begin{equation}
	\begin{aligned}
	&\lim\limits_{k\rightarrow\infty} \sum_{r=0}^{k} g\Big(\frac{r}{k}\Big)\int \dbinom{k}{r}\eta^r(1-\eta)^{k-r}f_{\mathbf{x}}(\mathbf{x})d\mathbf{x} \\
	=&\int \Bigg[ \lim\limits_{k\rightarrow \infty} \sum_{r=0}^{k} g\bigg(\frac{r}{k}\bigg) \dbinom{k}{r}\eta^r(1-\eta)^{k-r}\Bigg]f_{\mathbf{x}}(\mathbf{x})d\mathbf{x},
	\end{aligned}	
	\end{equation}	
	which according to Weierstrass' Approximation Theorem simplifies to 
	\begin{equation}
	\begin{aligned}
	&=	\int g(\eta)f_{\mathbf{x}}(\mathbf{x})d\mathbf{x} \\
	&=G(f_0,f_1).
	\end{aligned}
	\end{equation}
	Therefore
	\begin{equation}
	\lim\limits_{k\rightarrow \infty}\lim_{\substack{N\to \infty\\ k/N\to 0}} E\bigg[\Big( \hat{G}_{k,N}(\mathbf{X})-G(f_0,f_1)\Big)^2\bigg]=0.
	\end{equation}

	\section{Proof of Theorem \ref{thm:f_eta} }

	Starting with the augmented data set defined in Appendix \ref{sec:proof_thm1}, remember that $\Phi_k'(\mathbf{x}^*)$ is the sum of $k$ i.i.d. Bernoulli random variables. Therefore $\frac{1}{k}\Phi'_k(\mathbf{x}^*)$ represents the arithmetic mean of these output values, and 
	\begin{equation}
	\lim_{k\rightarrow \infty}\frac{1}{k}\Phi'_k(\mathbf{x}^*)=\lim_{k\rightarrow \infty}\frac{1}{k}\sum_{i: \mathbf{x}_i\in\mathcal{N}_k(\mathbf{x}^*)}^{}y_i'= \eta(\mathbf{x}^*).
	\label{eq:phi_to_eta}
	\end{equation}
	Now, since $\Phi_k'(\mathbf{x})$ must be an integer, its probability of equaling $r$ can be expressed as
	\begin{equation}
	P[\Phi_k'(\mathbf{x})=r]=P\Bigg[ r-1< \Phi_k'(\mathbf{x})\leq r\Bigg].
	\label{eq:phi_limit}
	\end{equation}
	 Taking the limit of this expression w.r.t. $k$, allows us to use (\ref{eq:phi_to_eta}) to form
	\begin{equation}
	\begin{aligned}
\lim\limits_{k\rightarrow \infty}P\Bigg[ r-1< \Phi_k'(\mathbf{x})\leq r\Bigg]&=\lim_{k\rightarrow \infty}P\bigg[\frac{r-1}{k}< \eta \leq \frac{r}{k}\bigg]\\
	&=\lim_{k\rightarrow \infty}F_\eta\Big(\frac{r}{k}\Big)-F_\eta\Big(\frac{r-1}{k}\Big).
	\end{aligned}
	\end{equation}
	Now if we multiply each side by $k$, the right hand side takes the form of Newton's difference quotient and can be simplified to the probability density function 
	\begin{equation}
	\begin{aligned}
	\lim\limits_{k\rightarrow \infty}kP[\Phi_k'(\mathbf{x})=r]&=\lim_{k\rightarrow \infty}\frac{F_\eta(\frac{r}{k})-F_\eta(\frac{r}{k}-\frac{1}{k})}{\frac{1}{k}}\\
	&=\lim_{k\rightarrow \infty}\frac{d}{d\eta}F_{\eta}\Big(\frac{r}{k}\Big)=\lim_{k\rightarrow \infty}f_{\eta}\Big(\frac{r}{k}\Big).
	\end{aligned}
	\end{equation}
	Since $\rho_{r,k,N}\rightarrow P[\Phi'_k(\mathbf{x})=r]$ as $N$ and $k$ approach infinity in a linked manner such that $k/N\rightarrow 0$ and $\frac{r}{k}\rightarrow \eta^*$
	\begin{equation}
	k\rho_{r,k,N}(\mathbf{X})\rightarrow f_{\eta}(\eta^*). \quad 
	\end{equation}

	\bibliographystyle{IEEEtran}
	\bibliography{IEEEabrv,References,ReferencesVB}

\begin{thebibliography}{10}
\providecommand{\url}[1]{#1}
\csname url@samestyle\endcsname
\providecommand{\newblock}{\relax}
\providecommand{\bibinfo}[2]{#2}
\providecommand{\BIBentrySTDinterwordspacing}{\spaceskip=0pt\relax}
\providecommand{\BIBentryALTinterwordstretchfactor}{4}
\providecommand{\BIBentryALTinterwordspacing}{\spaceskip=\fontdimen2\font plus
\BIBentryALTinterwordstretchfactor\fontdimen3\font minus
  \fontdimen4\font\relax}
\providecommand{\BIBforeignlanguage}[2]{{%
\expandafter\ifx\csname l@#1\endcsname\relax
\typeout{** WARNING: IEEEtran.bst: No hyphenation pattern has been}%
\typeout{** loaded for the language `#1'. Using the pattern for}%
\typeout{** the default language instead.}%
\else
\language=\csname l@#1\endcsname
\fi
#2}}
\providecommand{\BIBdecl}{\relax}
\BIBdecl

\bibitem{moreno2003kullback}
P.~J. Moreno, P.~P. Ho, and N.~Vasconcelos, ``A {K}ullback-{L}eibler divergence
  based kernel for {SVM} classification in multimedia applications,'' in
  \emph{Advances in neural information processing systems}, 2003, pp.
  1385--1392.

\bibitem{hamza2003image}
A.~B. Hamza and H.~Krim, ``Image registration and segmentation by maximizing
  the {J}ensen-{R}{\'e}nyi divergence,'' in \emph{Energy Minimization Methods
  in Computer Vision and Pattern Recognition}.\hskip 1em plus 0.5em minus
  0.4em\relax Springer, 2003, pp. 147--163.

\bibitem{hild2001blind}
K.~E. Hild, D.~Erdogmus, and J.~C. Principe, ``Blind source separation using
  {R}\'{e}nyi's mutual information,'' \emph{\textit{Signal Processing Letters,
  IEEE}}, vol.~8, no.~6, pp. 174--176, 2001.

\bibitem{banerjee2005clustering}
A.~Banerjee, S.~Merugu, I.~S. Dhillon, and J.~Ghosh, ``Clustering with
  {B}regman divergences,'' \emph{\textit{The Journal of Machine Learning
  Research}}, vol.~6, pp. 1705--1749, 2005.

\bibitem{ali1966general}
S.~Ali and S.~D. Silvey, ``A general class of coefficients of divergence of one
  distribution from another,'' \emph{\textit{Journal of the Royal Statistical
  Society. Series B (Methodological)}}, pp. 131--142, 1966.

\bibitem{nguyen2009surrogate}
X.~Nguyen, M.~J. Wainwright, and M.~I. Jordan, ``On surrogate loss functions
  and f-divergences,'' \emph{\textit{The Annals of Statistics}}, pp. 876--904,
  2009.

\bibitem{wu2016minimax}
Y.~Wu and P.~Yang, ``Minimax rates of entropy estimation on large alphabets via
  best polynomial approximation,'' \emph{IEEE Transactions on Information
  Theory}, vol.~62, no.~6, pp. 3702--3720, 2016.

\bibitem{jiao2015minimax}
J.~Jiao, K.~Venkat, Y.~Han, and T.~Weissman, ``Minimax estimation of
  functionals of discrete distributions,'' \emph{IEEE Transactions on
  Information Theory}, vol.~61, no.~5, pp. 2835--2885, 2015.

\bibitem{wu2015optimal}
Y.~Wu and P.~Yang, ``Optimal entropy estimation on large alphabets via best
  polynomial approximation,'' in \emph{IEEE International Symposium on
  Information Theory}.\hskip 1em plus 0.5em minus 0.4em\relax IEEE, 2015, pp.
  824--828.

\bibitem{valiant2011estimating}
G.~Valiant and P.~Valiant, ``Estimating the unseen: an n/log (n)-sample
  estimator for entropy and support size, shown optimal via new {CLT}s,'' in
  \emph{Proceedings of the forty-third annual ACM symposium on Theory of
  computing}.\hskip 1em plus 0.5em minus 0.4em\relax ACM, 2011, pp. 685--694.

\bibitem{paninski2003estimation}
L.~Paninski, ``Estimation of entropy and mutual information,'' \emph{Neural
  computation}, vol.~15, no.~6, pp. 1191--1253, 2003.

\bibitem{paninski2004estimating}
------, ``Estimating entropy on m bins given fewer than m samples,'' \emph{IEEE
  Transactions on Information Theory}, vol.~50, no.~9, pp. 2200--2203, 2004.

\bibitem{hero01}
A.~O. Hero, B.~Ma, O.~Michel, and J.~Gorman, ``Alpha-divergence for
  classification, indexing and retrieval,'' \emph{\textit{Communication and
  Signal Processing Laboratory, Technical Report CSPL-328, U. Mich}}, 2001.

\bibitem{ahmad1976nonparametric}
I.~Ahmad and P.-E. Lin, ``A nonparametric estimation of the entropy for
  absolutely continuous distributions (corresp.),'' \emph{\textit{IEEE
  Transactions on Information Theory}}, vol.~22, no.~3, pp. 372--375, 1976.

\bibitem{gyorfi1987density}
L.~Gy{\"o}rfi and E.~C. Van~der Meulen, ``Density-free convergence properties
  of various estimators of entropy,'' \emph{\textit{Computational Statistics \&
  Data Analysis}}, vol.~5, no.~4, pp. 425--436, 1987.

\bibitem{izenman1991review}
A.~J. Izenman, ``Review papers: recent developments in nonparametric density
  estimation,'' \emph{\textit{Journal of the American Statistical
  Association}}, vol.~86, no. 413, pp. 205--224, 1991.

\bibitem{kozachenko1987sample}
L.~Kozachenko and N.~N. Leonenko, ``Sample estimate of the entropy of a random
  vector,'' \emph{Problemy Peredachi Informatsii}, vol.~23, no.~2, pp. 9--16,
  1987.

\bibitem{hero1998robust}
A.~O. Hero~III and O.~Michel, ``Robust entropy estimation strategies based on
  edge weighted random graphs,'' in \emph{SPIE's International Symposium on
  Optical Science, Engineering, and Instrumentation}.\hskip 1em plus 0.5em
  minus 0.4em\relax International Society for Optics and Photonics, 1998, pp.
  250--261.

\bibitem{pal2010estimation}
D.~P{\'a}l, B.~P{\'o}czos, and C.~Szepesv{\'a}ri, ``Estimation of r{\'e}nyi
  entropy and mutual information based on generalized nearest-neighbor
  graphs,'' in \emph{Advances in Neural Information Processing Systems}, 2010,
  pp. 1849--1857.

\bibitem{berisha2014empirically}
V.~Berisha, A.~Wisler, A.~O. Hero~III, and A.~Spanias, ``Empirically estimable
  classification bounds based on a nonparametric divergence measure,''
  \emph{\textit{IEEE Transactions on Signal Processing}}, vol.~64, no.~3, pp.
  580--591, 2016.

\bibitem{kandasamy2015nonparametric}
K.~Kandasamy, A.~Krishnamurthy, B.~Poczos, L.~Wasserman, and J.~Robins,
  ``Nonparametric von {Mises} estimators for entropies, divergences and mutual
  informations,'' in \emph{Advances in Neural Information Processing Systems},
  2015, pp. 397--405.

\bibitem{silva2010information}
J.~Silva and S.~S. Narayanan, ``Information divergence estimation based on
  data-dependent partitions,'' \emph{Journal of Statistical Planning and
  Inference}, vol. 140, no.~11, pp. 3180--3198, 2010.

\bibitem{darbellay1999estimation}
G.~A. Darbellay, I.~Vajda \emph{et~al.}, ``Estimation of the information by an
  adaptive partitioning of the observation space,'' \emph{IEEE Transactions on
  Information Theory}, vol.~45, no.~4, pp. 1315--1321, 1999.

\bibitem{geman1982nonparametric}
S.~Geman and C.-R. Hwang, ``Nonparametric maximum likelihood estimation by the
  method of sieves,'' \emph{Annals of Statistics}, pp. 401--414, 1982.

\bibitem{grenander1981abstract}
U.~Grenander and G.~Ulf, ``Abstract inference,'' Tech. Rep., 1981.

\bibitem{moon2014multivariate}
K.~Moon and A.~Hero, ``Multivariate f-divergence estimation with confidence,''
  in \emph{Advances in Neural Information Processing Systems}, 2014, pp.
  2420--2428.

\bibitem{moon2016improving}
K.~R. Moon, K.~Sricharan, K.~Greenewald, and A.~O. Hero, ``Improving
  convergence of divergence functional ensemble estimators,'' in \emph{IEEE
  International Symposium on Information Theory}.\hskip 1em plus 0.5em minus
  0.4em\relax IEEE, 2016, pp. 1133--1137.

\bibitem{poczos2012nonparametric}
B.~P{\'o}czos, L.~Xiong, and J.~Schneider, ``Nonparametric divergence
  estimation with applications to machine learning on distributions,''
  \emph{arXiv preprint arXiv:1202.3758}, 2012.

\bibitem{nguyen2010estimating}
X.~Nguyen, M.~J. Wainwright, and M.~I. Jordan, ``Estimating divergence
  functionals and the likelihood ratio by convex risk minimization,''
  \emph{\textit{IEEE Transactions on Information Theory}}, vol.~56, no.~11, pp.
  5847--5861, 2010.

\bibitem{wang2009divergence}
Q.~Wang, S.~R. Kulkarni, and S.~Verd{\'u}, ``Divergence estimation for
  multidimensional densities via $ k $-nearest-neighbor distances,'' \emph{IEEE
  Transactions on Information Theory}, vol.~55, no.~5, pp. 2392--2405, 2009.

\bibitem{wang2009universal}
------, ``Universal estimation of information measures for analog sources,''
  \emph{Foundations and Trends in Communications and Information Theory},
  vol.~5, no.~3, pp. 265--353, 2009.

\bibitem{poczos2011estimation}
B.~P{\'o}czos and J.~G. Schneider, ``On the estimation of alpha-divergences,''
  in \emph{International Conference on Artificial Intelligence and Statistics},
  2011, pp. 609--617.

\bibitem{friedman1979multivariate}
J.~H. Friedman and L.~C. Rafsky, ``Multivariate generalizations of the
  {W}ald-{W}olfowitz and {S}mirnov two-sample tests,'' \emph{\textit{The Annals
  of Statistics}}, pp. 697--717, 1979.

\bibitem{henze1999multivariate}
N.~Henze, M.~D. Penrose \emph{et~al.}, ``On the multivariate runs test,''
  \emph{\textit{The Annals of Statistics}}, vol.~27, no.~1, pp. 290--298, 1999.

\bibitem{chernoff1952measure}
H.~Chernoff, ``A measure of asymptotic efficiency for tests of a hypothesis
  based on the sum of observations,'' \emph{\textit{The Annals of Mathematical
  Statistics}}, pp. 493--507, 1952.

\bibitem{bhattacharyya1946measure}
A.~Bhattacharyya, ``On a measure of divergence between two multinomial
  populations,'' \emph{\textit{Sankhy{\=a}: The Indian Journal of Statistics}},
  pp. 401--406, 1946.

\bibitem{kailath1967divergence}
T.~Kailath, ``The divergence and {B}hattacharyya distance measures in signal
  selection,'' \emph{\textit{Communication Technology, IEEE Transactions on}},
  vol.~15, no.~1, pp. 52--60, 1967.

\bibitem{hashlamoun1994tight}
W.~A. Hashlamoun, P.~K. Varshney, and V.~Samarasooriya, ``A tight upper bound
  on the {B}ayesian probability of error,'' \emph{\textit{IEEE Transactions on
  Pattern Analysis and Machine Intelligence}}, vol.~16, no.~2, pp. 220--224,
  1994.

\bibitem{avi1996arbitrarily}
H.~Avi-Itzhak and T.~Diep, ``Arbitrarily tight upper and lower bounds on the
  {B}ayesian probability of error,'' \emph{\textit{IEEE Transactions on Pattern
  Analysis and Machine Intelligence}}, vol.~18, no.~1, pp. 89--91, 1996.

\bibitem{berisha2014empirical}
V.~Berisha and A.~O. Hero, ``Empirical non-parametric estimation of the
  {F}isher information,'' \emph{IEEE Signal Processing Letters}, vol.~22,
  no.~7, pp. 988--992, 2015.

\bibitem{leblanc2012estimating}
A.~Leblanc, ``On estimating distribution functions using {B}ernstein
  polynomials,'' \emph{Annals of the Institute of Statistical Mathematics},
  vol.~64, no.~5, pp. 919--943, 2012.

\bibitem{turnbull2014unimodal}
B.~C. Turnbull and S.~K. Ghosh, ``Unimodal density estimation using {B}ernstein
  polynomials,'' \emph{Computational Statistics \& Data Analysis}, vol.~72, pp.
  13--29, 2014.

\bibitem{ghosal2001convergence}
S.~Ghosal, ``Convergence rates for density estimation with {B}ernstein
  polynomials,'' \emph{Annals of Statistics}, pp. 1264--1280, 2001.

\bibitem{igarashi2014improving}
G.~Igarashi and Y.~Kakizawa, ``On improving convergence rate of {B}ernstein
  polynomial density estimator,'' \emph{Journal of Nonparametric Statistics},
  vol.~26, no.~1, pp. 61--84, 2014.

\bibitem{tenbusch1994two}
A.~Tenbusch, ``Two-dimensional {B}ernstein polynomial density estimators,''
  \emph{Metrika}, vol.~41, no.~1, pp. 233--253, 1994.

\bibitem{babu2006smooth}
G.~J. Babu and Y.~P. Chaubey, ``Smooth estimation of a distribution and density
  function on a hypercube using {B}ernstein polynomials for dependent random
  vectors,'' \emph{Statistics \& probability letters}, vol.~76, no.~9, pp.
  959--969, 2006.

\bibitem{lorentz2012bernstein}
G.~G. Lorentz, \emph{{B}ernstein polynomials}.\hskip 1em plus 0.5em minus
  0.4em\relax American Mathematical Soc., 2012.

\bibitem{bernstein1912démo}
S.~Bernstein, ``D\'{e}monstration du th\'{e}or\`{e}me de {Weierstrass}
  fond\'{e}e sur le calcul des probabilit\'{e}s,'' \emph{Comm. Soc. Math.
  Kharkow, Ser}, vol.~2, no.~13, pp. 49--194, 1912.

\bibitem{gubner2006probability}
J.~A. Gubner, \emph{Probability and random processes for electrical and
  computer engineers}.\hskip 1em plus 0.5em minus 0.4em\relax Cambridge
  University Press, 2006.

\bibitem{farouki2000legendre}
R.~T. Farouki, ``Legendre--{B}ernstein basis transformations,'' \emph{Journal
  of Computational and Applied Mathematics}, vol. 119, no.~1, pp. 145--160,
  2000.

\bibitem{efron2004least}
B.~Efron, T.~Hastie, I.~Johnstone, R.~Tibshirani \emph{et~al.}, ``Least angle
  regression,'' \emph{The Annals of statistics}, vol.~32, no.~2, pp. 407--499,
  2004.

\bibitem{arya1998optimal}
S.~Arya, D.~M. Mount, N.~S. Netanyahu, R.~Silverman, and A.~Y. Wu, ``An optimal
  algorithm for approximate nearest neighbor searching fixed dimensions,''
  \emph{Journal of the ACM (JACM)}, vol.~45, no.~6, pp. 891--923, 1998.

\bibitem{sutherland2012kernels}
D.~J. Sutherland, L.~Xiong, B.~P{\'o}czos, and J.~Schneider, ``Kernels on
  sample sets via nonparametric divergence estimates,'' \emph{arXiv preprint
  arXiv:1202.0302}, 2012.

\bibitem{szabo14information}
Z.~Szab{\'o}, ``Information theoretical estimators toolbox,'' \emph{Journal of
  Machine Learning Research}, vol.~15, pp. 283--287, 2014.

\bibitem{kullback1997information}
S.~Kullback, \emph{Information theory and statistics}.\hskip 1em plus 0.5em
  minus 0.4em\relax Courier Corporation, 1997.

\bibitem{wisler2016multiclass}
A.~Wisler, V.~Berisha, K.~Ramamurthy, D.~Wei, and A.~Spanias,
  ``Emperically-estimable multi-class performance bounds,'' in \emph{IEEE
  International Conference on Acoustics, Speech and Signal Processing
  (ICASSP)}.\hskip 1em plus 0.5em minus 0.4em\relax IEEE, 2016.

\bibitem{fukunaga1990introduction}
K.~Fukunaga, \emph{Introduction to statistical pattern recognition}.\hskip 1em
  plus 0.5em minus 0.4em\relax Academic press, 1990.

\bibitem{devroye2013probabilistic}
L.~Devroye, L.~Gy{\"o}rfi, and G.~Lugosi, \emph{A probabilistic theory of
  pattern recognition}.\hskip 1em plus 0.5em minus 0.4em\relax Springer Science
  \& Business Media, 2013, vol.~31.

\bibitem{van2000asymptotic}
A.~W. Van~der Vaart, \emph{Asymptotic statistics}.\hskip 1em plus 0.5em minus
  0.4em\relax Cambridge university press, 2000, vol.~3.

\end{thebibliography}

\end{document}